\documentclass[a4paper,12pt]{article}
\usepackage{epsfig}
\usepackage{graphicx}
\usepackage{subfigure}
\font\mybb=msbm10 at 11pt

\def\bb#1{\hbox{\mybb#1}}
\def\bR {\bb{R}}

\def\ie{{\it i.e.,}}
\newcommand{\be}{\begin{equation}}
\newcommand{\ee}{\end{equation}}
\newcommand{\bea}{\begin{eqnarray}}
\newcommand{\eea}{\end{eqnarray}}
\newcommand{\bml}{\begin{mathletters}}
\newcommand{\eml}{\end{mathletters}}

\newcommand{\fr}{\frac}

\newcommand{\al}{\alpha}

\newcommand{\pr}{\partial}

\newcommand{\acc}{\\[3mm]}
\newcommand{\dl}{\delta}

\newcommand{\I}{{\it I}}
\newcommand{\Ic}{{\cal I}}

\def\U {{\rlap{\kern 1.2mm \vrule height 7pt depth 0pt} \rm 1}}
\begin{document}


\renewcommand{\thefootnote}{\fnsymbol{footnote}}
\title{Mass terms in the Skyrme Model}
\author{Vladimir B. Kopeliovich\thanks{email: kopelio@al20.inr.troitsk.ru},\,
\\
Institute for Nuclear Research, Moscow 117312, Russia
\acc
\acc
Bernard Piette\thanks{email: b.m.a.g.piette@durham.ac.uk}\,
and
Wojtek J. Zakrzewski\thanks{email: w.j.zakrzewski@durham.ac.uk}
\\
Department of Mathematical Sciences, University of Durham, \\
Durham DH1 3LE, UK\\
}
\date{\today}

\maketitle

\setlength{\footnotesep}{0.5\footnotesep}
\begin{abstract}
We consider various forms of the mass term that can be used in the 
Skyrme model
and their implications on the properties of baryonic states.
We show that, with an appropriate choice for the mass term, without changing 
the asymptotic behaviour of the profile functions at large $r$, we can 
considerably reduce or increase the mass term's contribution
 to the classical mass of the solitons. We find that multibaryon 
configurations can be classically bound at large baryon numbers 
for some choices of this mass term.
\end{abstract}

\section{Introduction}
The Skyrme model has enjoyed a lot of interest ever since it was realised 
that, although it is a nonlinear theory of pions, 
 it is also an effective theory of low energy nucleon interactions.
In fact, it may also provide a new approach to nuclear physics; as 
the lowest states of the model, corresponding 
to higher baryon numbers, are expected to provide a classical 
description of nuclei. In the Skyrme model approach the baryon number is 
identified with the soliton number.

Multiskyrmions are the stationary points of the
static energy functional which, in natural units of the model, 
$3\pi\sp2 F_\pi/e$, is given  by
\be
E=\fr{1}{12\pi^2}\int_{R^3}\left\{-\fr{1}{2}\mbox{Tr}\left(\pr_iU\,
U^{-1}\right)^2+\fr{1}{16}\mbox{Tr}\left[\pr_iU\,
U^{-1},\;\pr_j U\, U^{-1}\right]^2\right\}d^3\vec{x}.
\label{gene}
\ee
where $U(\vec{x})\in SU(2)$ and $x$ is in units of $2/(F_\pi e)$.

Most of the phenomenological applications of the Skyrme model, especially to 
the study of the nucleon or hyperon
properties, included also the pion (or kaon, or $D$-meson) mass term in the 
Lagrangian chosen  in the simplest possible form (see eg
Adkins and Nappi \cite{AN}). 
In particular, the kaon mass 
term has to be added to describe the mass splittings
within the $SU(3)$ multiplets of baryons: octet, decuplet, antidecuplet, etc 
\cite{G}. 
However, the role of the mass terms in multiskyrmion configurations, 
especially at large baryon 
numbers, has not been investigated in much detail;
 the theoretical work performed so far has involved mostly the Skyrme model
in which pions are massles ({\it i.e.} given by the Lagrangian above).
It is only very recently that some attention has been paid also 
to the effects associated with the pion mass for large $B$ configurations 
\cite{vk,BS};  one of the
effects being the exponential localization of multiskyrmions. In particular, 
it was stressed that
the contribution of the mass term can change the binding properties of large 
$B$ classical configurations and,  in particu1lar, their decay properties into 
configurations with smaller $B$-numbers \cite{BS}.

In most of these approaches the pion mass term has been introduced via the 
addition to
(\ref{gene}) of the following term
\be
\fr{1}{12\pi^2}\int_{R^3} \,m\sp2\, \mbox{Tr}(1-U)\,d^3\vec{x}.
\label{ma}
\ee
where $m$ is related to physical pion mass $\mu_\pi \simeq 138\,Mev$ by the 
relation 
$m=2\mu_\pi/(F_\pi e)$,
where $F_\pi\simeq 186 \,Mev$ is the pion decay constant, taken usually 
from the experiment, 
and $e$ is a
Skyrme constant\footnote{In \cite{AN} the masses of the nucleon and the 
$\Delta (1232)$ isobar were fitted using an $SU(2)$ quantization procedure 
and, as a result, the authors obtained $F_\pi=108\,Mev,\; e=4.84$, but these 
values 
did not allow to describe the mass splittings within $SU(3)$ multiplets of 
baryons. The approach  of \cite{AN} has been revised and another set of 
parameters is widely accepted now. The baryon mass splittings are described 
with the experimental 
values of $F_\pi,\;F_K$ and $e\simeq 4.1$. For these parameters the absolute 
values of the baryon masses are not fitted because they are 
controlled by the loop corrections, or the so-called Casimir energy which, 
for the baryon number $1$, was estimated in \cite{M}.}.
The appearance of the mass term in effective field theories was discussed, 
e.g. in \cite{GL}.
Although, the effects associated with the pion mass are small for the small 
values
of this mass, they  increase if either the baryon number or the pion mass 
are larger.  For massless pions all the known minimal energy multiskyrmion
configurations have a shell-like structures. These field configurations were
obtained in both numerical simulations and in studies involving the 
so-called
`rational map ansatz'. In the rational map ansatz one approximates the full 
multiskyrmion field by assuming that its angular dependence is approximately
described by a rational map between Riemann spheres. This approximation was, 
first of all,
 shown to be very good in a theory with massless pions and it was later 
extended 
 also to massive pions - where the agreement was again shown to be very good.
 
 Given these observations it is extremely important to have the right 
(correct) mass
 term. The problem, however, is that the mass term is very non-unique and the 
expression
 (\ref{ma}) is only one of many that can be used. Indeed, the origin of 
the chiral symmetry conserving and chiral symmetry breaking, or the mass terms 
considered in \cite{GL}, may have a very different nature.
 So we have decided to reexamine this issue further and to look at mass terms 
other than (\ref{ma}) and see what effects they have on the properties of 
multiskyrmion configurations.
 
In the next section we discuss various choices of the mass term, pointing out 
what is fixed and what can be changed, and in the following sections we look 
at some simple examples of such mass terms. Expressions for the static energy 
of skyrmions
and some definitions necessary for the description of multiskyrmions within the
rational map approximation are presented in sections 3 and 4. Our numerical 
results 
are presented in section 5 and the analytical discussion useful to 
establish asymptotic behaviour is presented in section 6. We finish with a 
short section discussing our conclusions and ideas for further work.
\section{Mass Terms}
To consider the mass term we first note that the pion fields 
${\vec {\pi}}$ = $(\pi_1,\pi_2,\pi_3)$
 are given by $U=\sigma+i\vec{\pi}\cdot\vec{\tau}$, where $\vec{\tau}$ denotes 
the triplet 
 of Pauli matrices and $\sigma$ is determined by the constraint 
$\sigma\sp2+\vec{\pi}\cdot\vec{\pi}=1 $.
 
 Then the square of the pion mass is the coefficient of the expansion of the 
mass term
 in powers of $\vec{\pi}\cdot\vec{\pi}$; in fact it is the coefficient
 of the lowest term {\it i.e.} $\vec{\pi}\cdot\vec{\pi}$ in this expansion. 
 In the case above we have 
\be 
 m\sp2\mbox{Tr}(1-U)\,=\,m\sp2 \, 2(1 - \sigma)\,\sim  m\sp2 \,
 \vec{\pi}\cdot\vec{\pi}\,+\,...,
 \label{ma1}
\ee
 where $+...$ stands for further powers of $\vec{\pi}\cdot\vec{\pi}$ to be 
interpreted
 as pion interaction terms. So the mass of the pion field is proportional to 
$m$, since the
 canonical mass term in the lagrangian is 
$- \mu_\pi\sp2 \vec{\pi}\cdot\vec{\pi}/2$.
 
 However, (\ref{ma}) is not the only term we can use as the pion mass term. It 
is clear that we can 
 multiply $(1-U)$ in (\ref{ma}) by any function of $U$ which in the limit 
 $U\rightarrow 1$ reduces to 1. Thus we could multiply it by, say, ${(U+1)\over 2}$!
 
 In fact, a little thought shows that, instead of $U$ in (\ref{ma}), we can 
take
 \be 
 \int_{-\infty}\sp{\infty} \,g(p) \, U\sp{p} \, dp
 \label{maone}
 \ee
 where
 \be
 \int_{-\infty}\sp{\infty} \,g(p)  \, dp\,=\, 1\,\quad \mbox{and}\quad 
 \int_{-\infty}\sp{\infty} \,g(p)p\sp2 
  \, dp\,=\, 1.
  \label{test}
  \ee
The usual choice then corresponds to
\be 
g(p)\,=\, \delta(p-1).
\ee

As the second condition in (\ref{test}) can be eliminated by redefining 
the coefficient $m\sp2$ 
in (\ref{ma}) we see that a more general mass term is given by
\be
\fr{1}{12\pi^2}Am\sp2\,\int_{R^3} \, \mbox{Tr}\left[1-\int_{-\infty}\sp{\infty}
\,g(p) \, U\sp{p} \, dp
\right]\,d^3\vec{x},
\label{matwo}
\ee
where 
\be
A\sp{-1}\,=\,\int_{-\infty}\sp{\infty} \,g(p)p\sp2 
  \, dp.
  \label{aa}
\ee
\section{$B$=1 Skyrmion}
Consider first the case of one Skyrmion. The single Skyrmion has the hedgehog 
form
\be
U\,=\,\mbox{exp}(if(r)\hat{\vec{r}}\cdot \vec{\tau}),
\label{hedgehog}
\ee
where $\hat{\vec{r}}$ is the unit vector
in the $\vec{r}$ direction and $f(r)$ is the radial profile function 
which is required to satisfy the 
boundary conditions $f(0)= \pi$ and $f(\infty)=0.$

Putting (\ref{hedgehog}) into the  energy functional we find that the energy 
of the field is given by
\bea
E\,\,&=&\, \fr{1}{3\pi}\int_0\sp{\infty}\Big(r\sp2\dot f\sp2\,
   +\,2(\dot f\sp2+1) \sin\sp2f\,+\,
\fr{\sin\sp4f}{r\sp2}\,\nonumber\\
   &+&\,2Am\sp2r\sp2\left[1-\int_{-\infty}\sp{\infty}\,g(k)\,
 \cos(kf)\,dk\right]\Big)dr,
\label{genen}
\eea
where $A$ is given by (\ref{aa}).

Thus, for the minimal field, $f(r)$ satisfies the equation

\bea
\ddot f[r\sp2+2\sin\sp2f]\,&+&\,2r\dot f\,+\,2(\dot f\sp2-1)\sin f\cos f\,-\,
\frac{2\sin\sp3f\cos f}{r\sp2}\nonumber\\
&-&\,m\sp2r\sp2\,\frac{\int_{-\infty}\sp{\infty}\,g(k)\,\sin(kf)\,k\,dk}
{\int_{-\infty}\sp{\infty}\,g(k)\,k\sp2\,dk}\,=\,0.
\label{eq}
\eea

We have investigated several classes of such functions:
\begin{itemize}
\item $g(k)=\delta(k-p)$ for several values of $p$.

The cases of even or odd integer values for $p$ have also been investigated  
analytically.
Moreover, it is easy to notice that the mass term for $p > 1$ is smaller than 
that for 
$p=1$,
since $ (1 - \cos (pf))/p^2 = 2 \sin^2(pf/2)/p^2 $ and 
$\sin^2 (pf/2)/p^2 < \sin^2(f/2)$ for $p>1$.

\item $g$ given by a Gaussian centered around $p=1$, or around $p=0$.

\end{itemize}

In the latter two cases we have taken
\be \label{gauss}
g(p)=\frac{\sqrt{\lambda}}{\sqrt{\pi}}\mbox{exp}\left(-\lambda(p-1)\sp2\right)
   \quad \mbox{and}\quad
g(p)=\frac{\sqrt{\lambda}}{\sqrt{\pi}}\mbox{exp}\left(-\lambda p\sp2\right).
\ee

The corresponding expressions for the energy are given by
\bea
E\,\,&=&\, \fr{1}{3\pi}\int_0\sp{\infty}\Biggl\{r\sp2\dot f\sp2\,
           +\,2(\dot f\sp2+1)\sin\sp2f\,
           +\,\fr{\sin\sp4f}{r\sp2}\,\nonumber\\
           &+&\,2m\sp2r\sp2\fr{1}{\fr{2}{\sqrt{\pi\lambda}}+1+\fr{1}{2\lambda}}
\left[1-\cos f\, \mbox{exp}(-\fr{f\sp2}{4\lambda})\right]\Biggr\}dr,
\label{genena}
\eea

and
\bea
E\,\,&=&\, \fr{1}{3\pi}\int_0\sp{\infty}\Biggl\{r\sp2\dot f\sp2\,
         +\,2(\dot f\sp2+1)\sin\sp2f\,
         +\,\fr{\sin\sp4f}{r\sp2}\,\nonumber\\
         &+&\,4m\sp2r\sp2 \lambda \left[1-\mbox{exp}(-\fr{f\sp2}{4\lambda})
                \right]\Biggr\} dr,
\label{genenb},
\eea
respectively.
\section{Multiskyrmions}
For multiskyrmion fields we use the rational map ansatz of 
Houghton et al. \cite{HMS}.
The ansatz involves the introduction of the spherical coordinates in $\bR^3$, 
so that a point ${\bf x}\in\bR^3$ is given  by a pair $(r,\xi)$, where
$r=\vert{\bf x}\vert$ is the  distance from the origin, and $\xi$ is a
Riemann sphere coordinate giving the point on the unit two-sphere
which intersects the half-line through the origin and the point 
${\bf x}$, \ie\ $\xi=\tan({\theta\over 2})e\sp{i\varphi}$, where $\theta$ and 
$\varphi$ are the usual spherical coordinates on the unit sphere.

Then one observes\cite{IPZ} that a general $SU(2)$ matrix, $U$, can always be 
written in the form
\be
U=\exp(if(2P-\I))
\label{rma1}
\ee
where $f$ is real and $P$ is a $2\times 2$ hermitian projector 
\ie\ $P=P^2=P^\dagger.$ The rational map ansatz assumes that the
Skyrme field has the above form and, in addition, that $f$ depends only
on the radial coordinate, \ie\  $f=f(r)$, and that the projector depends only
on the angular coordinates,\ie\  $P(\xi,\bar\xi).$

The projector is then taken in the form
\be P=\frac{{\bf f}\otimes {\bf f}^\dagger}{\vert {\bf f}\vert^2} 
\label{rma2}
\ee
where ${\bf f}(\xi)$ is a 2-component vector, each
entry of which is a degree $k$ polynomial in $\xi.$ 
Incidentally, given
 the projective nature of ${\bf f}$ one can also use the parametrization
${\bf f}=(1,R)^t$, where
$R(\xi)$ is the ratio of $R={f_1\over f_2}$. 

For $B=1$ this ansatz reproduces the one Skyrmion field configuration 
discussed in the last section, while for 
 $B>1$ the ansatz (\ref{rma1}) is not compatible with the equations which 
come from (\ref{gene}), so the ansatz cannot produce any
exact multi-skyrmion configurations. However, 
as was shown in many papers, see eg: \cite{HMS}, \cite{BS1}, 
it gives
approximate field configurations which turn out to be very close to
the numerically computed minimal energy states. To do this one selects a 
specific
map ${\bf f}$ and puts it into the Skyrme energy functional (\ref{gene}).
Performing the integration over the angular coordinates results in a
one-dimensional energy functional for $f(r)$ which has then to be solved
numerically. 

Hence, if we  write ${\bf f}=(1,R)^t$, then the Skyrme energy is
\bea
E&=&\frac{1}{3\pi}\int \bigg( r^2f'^2+2B(f'^2+1)\sin^2 f+\Ic\frac{\sin^4
f}{r^2}\bigg) \ dr\,\nonumber\\
&+&\,2Am\sp2r\sp2\left[1-\int_{-\infty}\sp{\infty}\,g(k)\,
 \cos(kf)\,dk\right]\Big)dr,
\label{rmaenergy}
\eea
where $\Ic$ denotes the integral 
\be \Ic=\frac{1}{4\pi}\int \bigg(
\frac{1+\vert \xi\vert^2}{1+\vert R\vert^2}
\bigg\vert\frac{dR}{d\xi}\bigg\vert\bigg)^4 \frac{2i \  d\xi  d\bar \xi
}{(1+\vert \xi\vert^2)^2}\,.
\label{i}
\ee 
The values of $\Ic$ have already been calculated; in what follows we take our 
values from \cite{BS1}. The equation for the profile function $f$ is very 
similar to the equation (\ref{eq}) of the last section except that the 
coefficients of terms involving $\sin f\cos f$ are multiplied 
by $B$ in the first term and $\Ic$ in the second.
\section{Numerical results}
We have looked at the values of various quantities for different choices 
of $p$ (taking $g(k)=\delta(k-p)$), and also at some Gaussians.
The Gaussian cases were not particularly illuminating so here we discuss
only the cases of fixed values of $p$. 

Note that when $p\rightarrow0$ we have a nontrivial
contribution of the mass term. This involves taking the limit 
$p\rightarrow0$ of the expression in (\ref{rmaenergy}) and then its last 
line becomes
$m^2r^2f^2$. We can also consider the limit when $m\rightarrow0$ which 
corresponds to the massless Skyrme model.

\begin{figure}[h]
\unitlength1cm \hfil
\begin{picture}(15,9)
 \epsfxsize=8cm \put(0,0){\epsffile{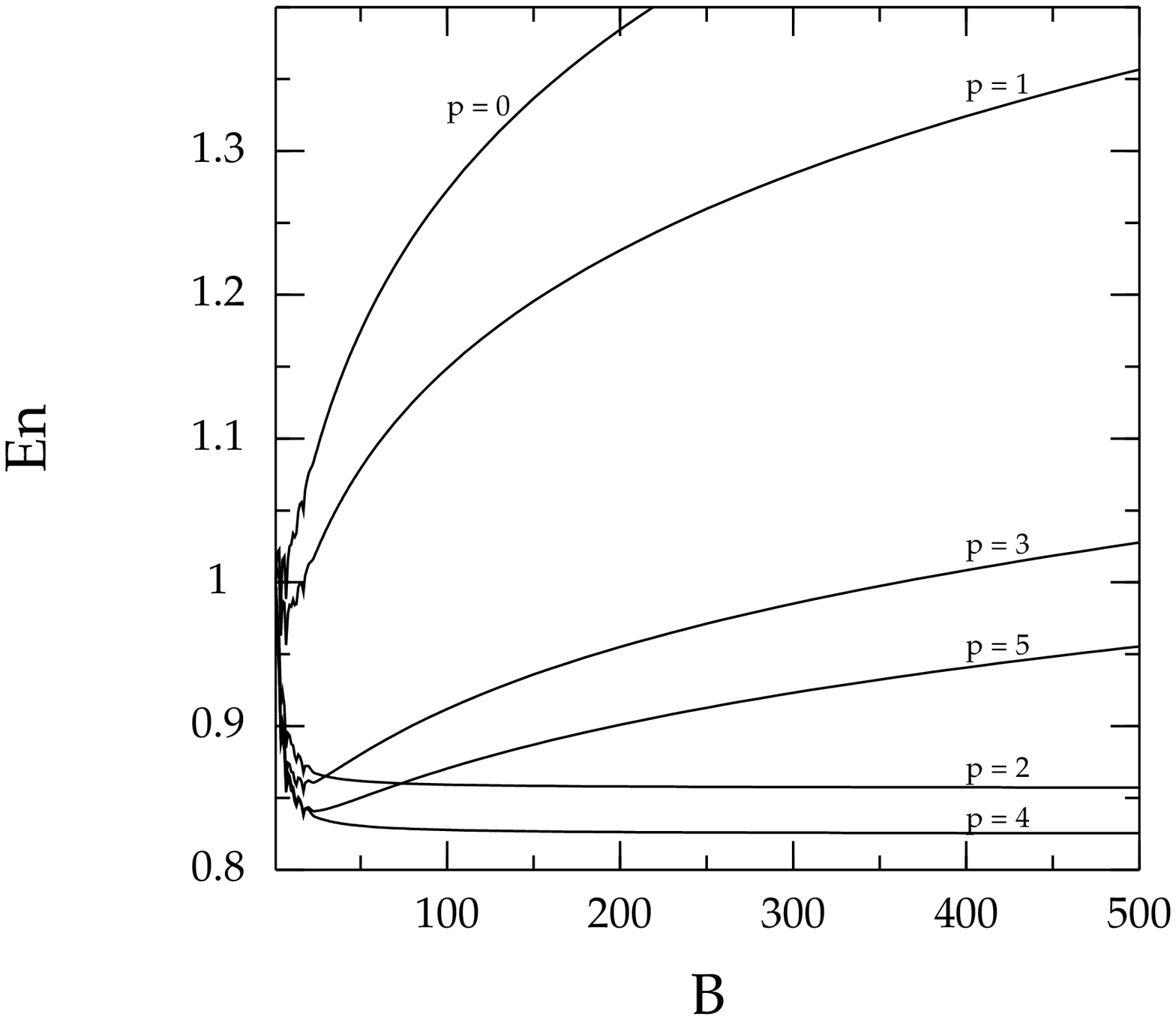}}
 \epsfxsize=8cm \put(7,0){\epsffile{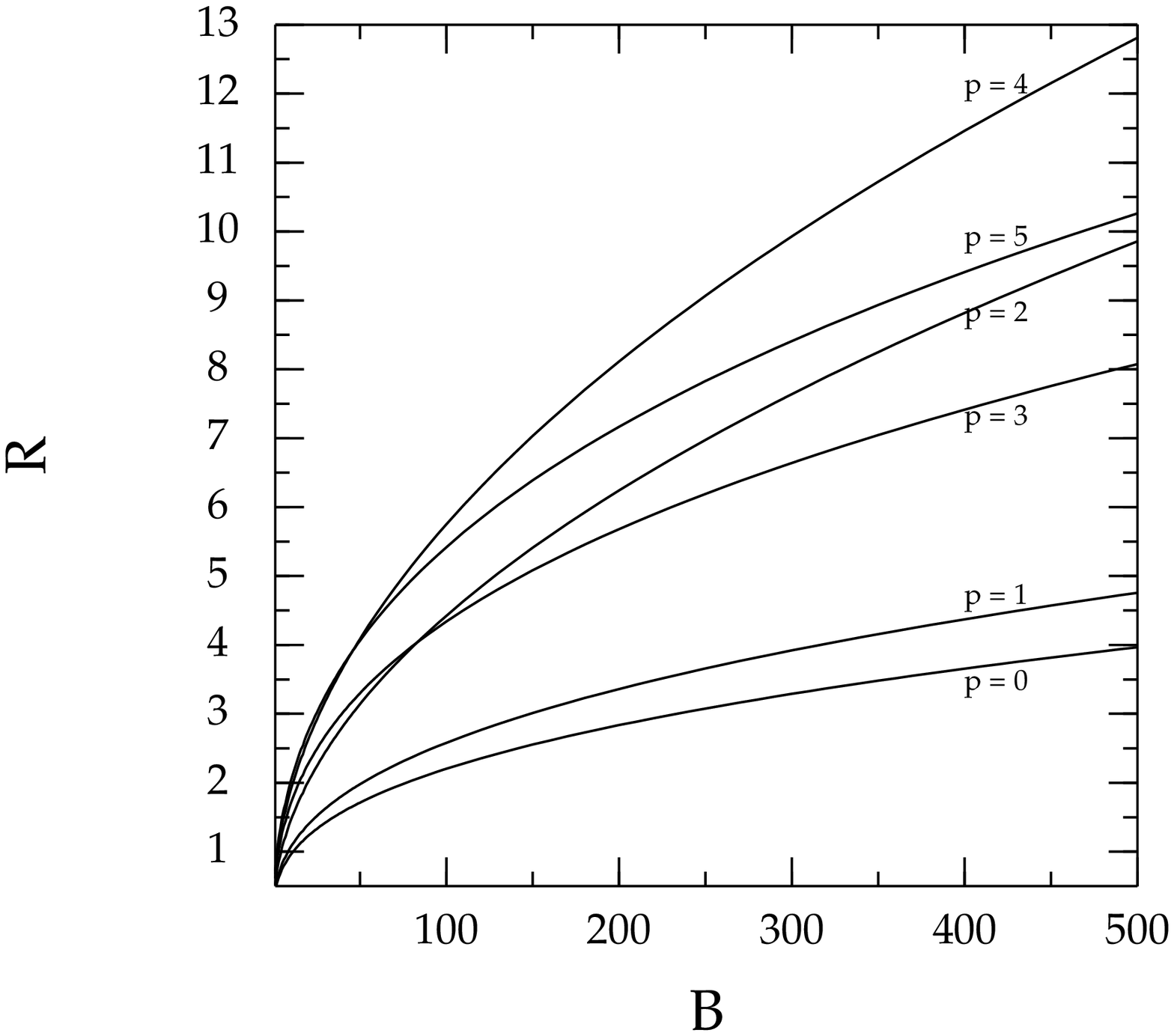}}
\put(3,0){a}
\put(10.0,0){b}
\end{picture}
\caption{\label{figch}
Normalized energy (\ref{NorEn}) (a) and radius (\ref{RE}) (b) of multiskyrmion 
configurations for $m=4.131$, $E(1)=2.056$}
\end{figure}

\begin{figure}[h]
\unitlength1cm \hfil
\begin{picture}(15,9)
 \epsfxsize=8cm \put(0,0){\epsffile{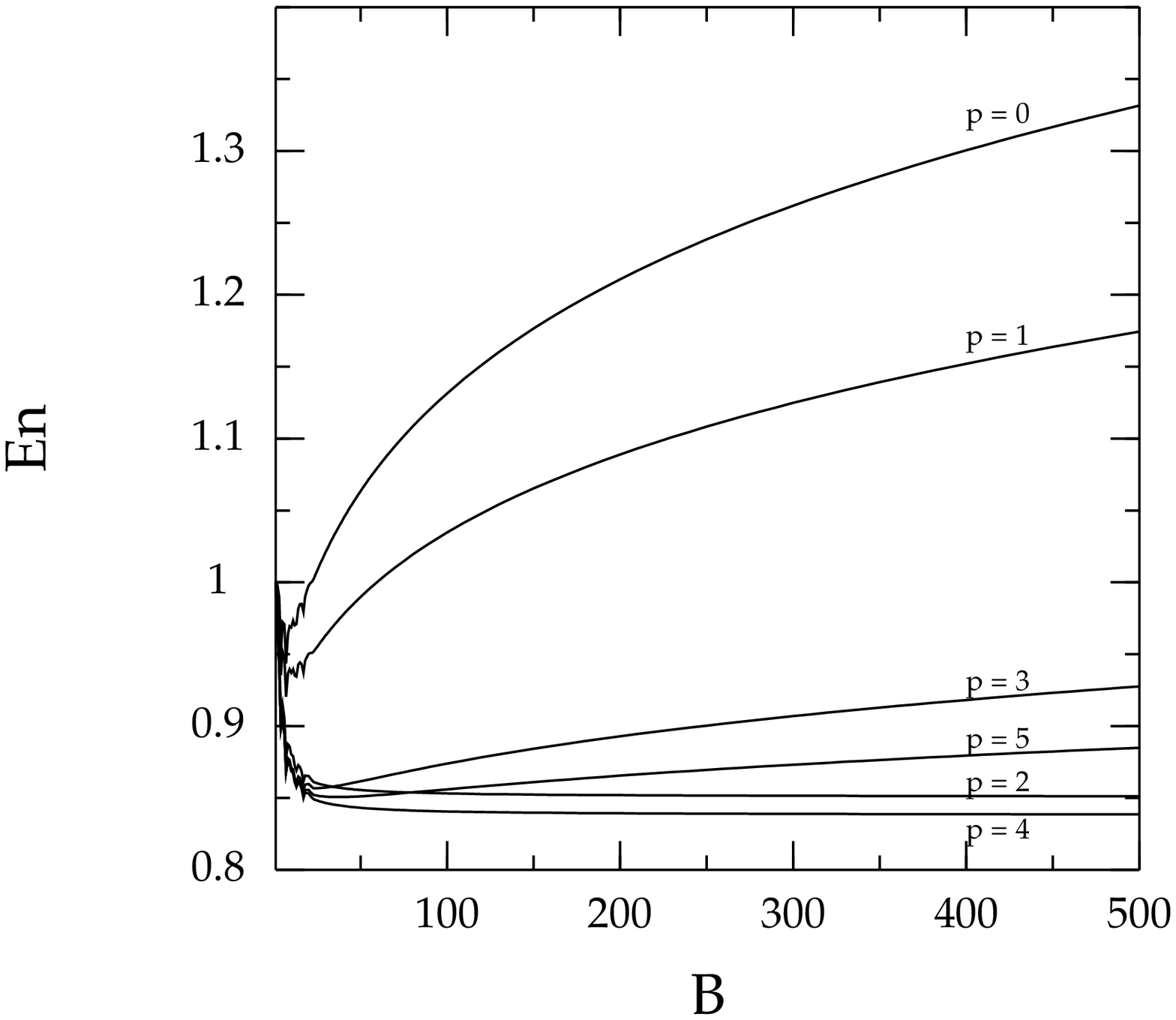}}
 \epsfxsize=8cm \put(7,0){\epsffile{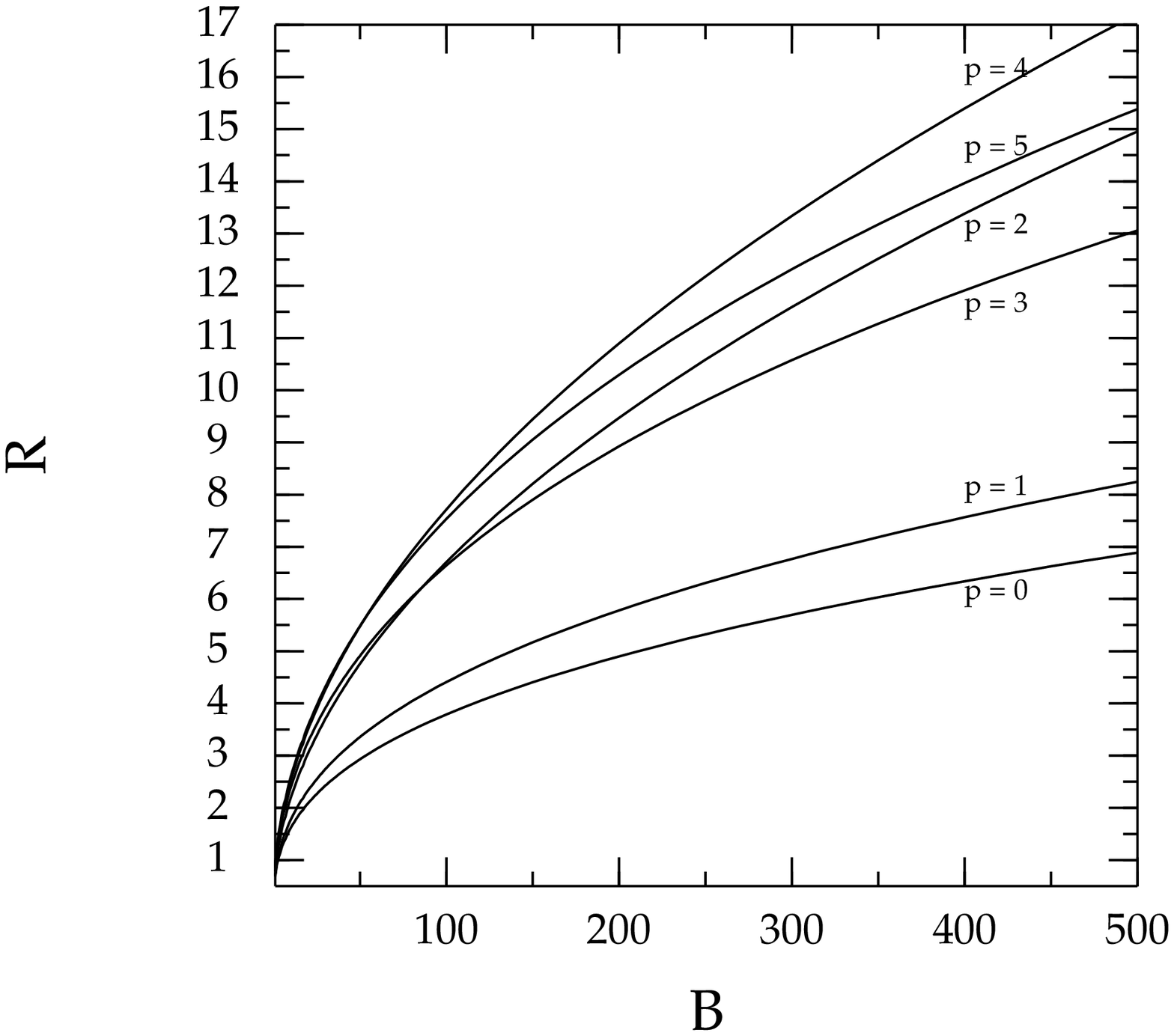}}
\put(3.0,0){a}
\put(10.0,0){b}
\end{picture}
\caption{\label{figst}
Normalized energy (\ref{NorEn}) (a) and radius (\ref{RE}) (b) of multiskyrmion 
configurations for $m=1.300$, $E(1)=1.486$.}
\end{figure}

\begin{figure}[h]
\unitlength1cm \hfil
\begin{picture}(15,9)
 \epsfxsize=8cm \put(0,0){\epsffile{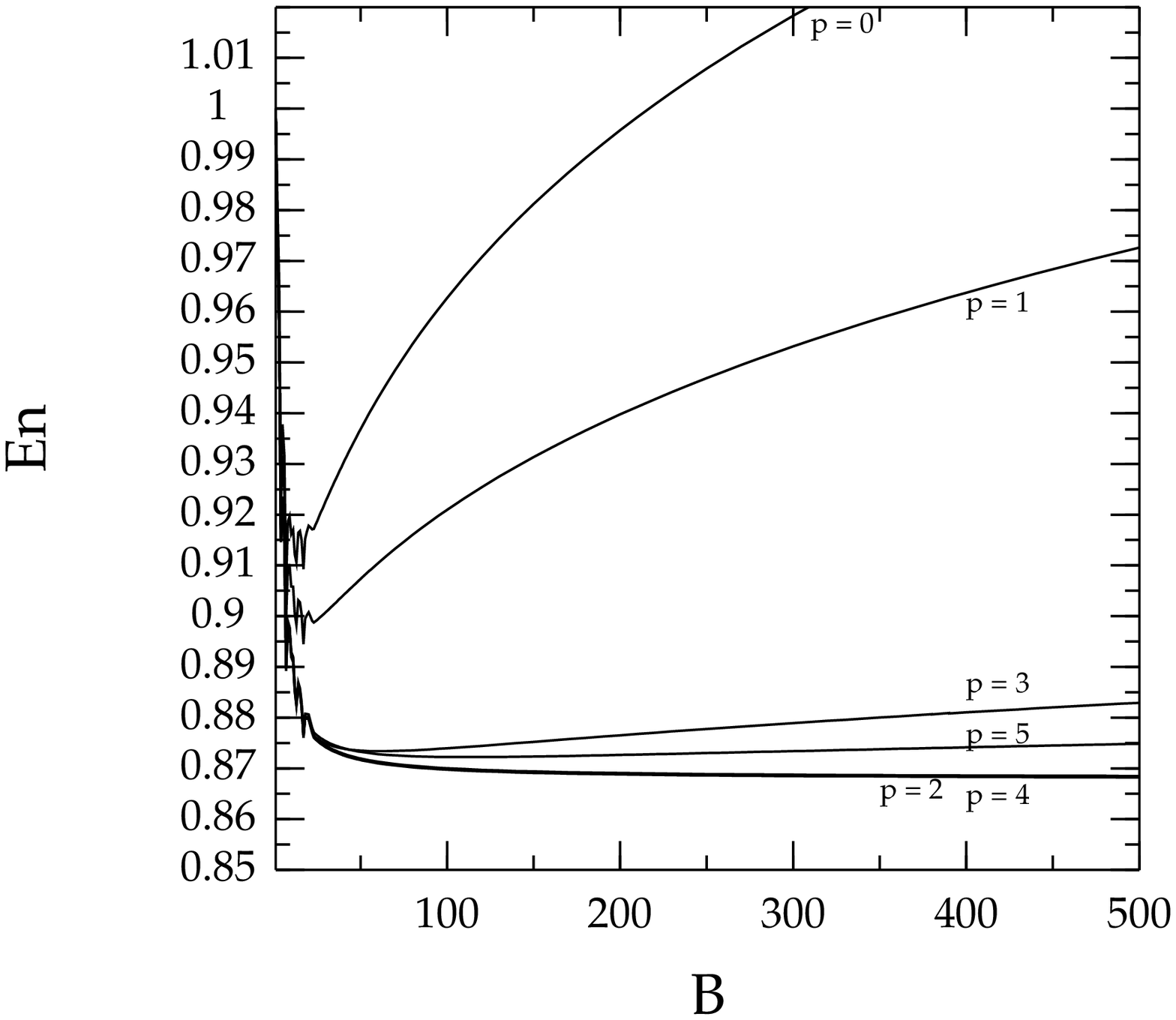}}
 \epsfxsize=8cm \put(7,0){\epsffile{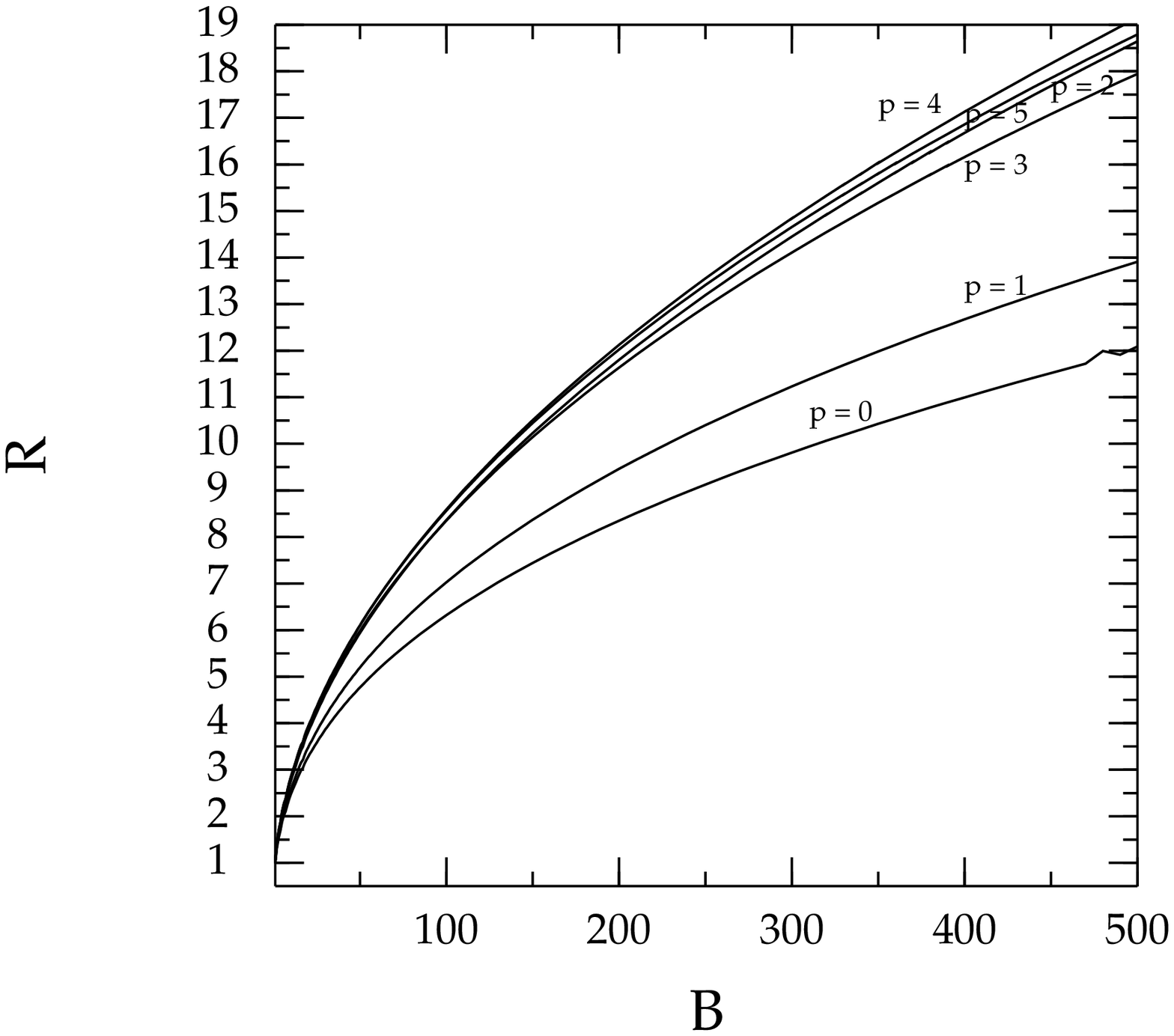}}
\put(3.0,0){a}
\put(10.0,0){b}
\end{picture}
\caption{\label{figpi}
Normalized energy (\ref{NorEn}) (a) and radius (\ref{RE}) (b) of multiskyrmion 
configurations for $m=0.362$, $E(1)=1.274$.}
\end{figure}

We present our numerical results in figures \ref{figch}-\ref{figpi}
in which we plot the normalized energy 
\begin{equation}
En = {E(B)\over B\, E(1)}
\label{NorEn}
\end{equation}
and the shell radius, as a function of $B$,
for several values of the mass \footnote{In our numerical calculations we
take for pions $m=m_\pi = 0.36192$, for kaons $m_{st}=1.29996$ and for the 
charmed mass scale $m_{ch}=4.130964$. In what follows, in the text and in 
the captions, we refer to those values as $m=0.362,\; 1.300$ and 
$4.131$ respectively .} and for $p$ from $0$ to $5$.

The normalized energy (\ref{NorEn}) is a dimensionless energy which describes 
the binding of the configuration by comparing it to that of the $B=1$ solution.
Note that when $En > 1$ the $B$ multiskyrmion configuration has an energy 
larger than the energy of $B$ single skyrmions thus showing that this 
configuration is unstable. The jagged curve near the origin is caused by 
the value
of $\Ic$ which varies a lot when $B$ is small. When $B>22$, we have taken 
$\Ic=1.28 B^2$ (see \cite{BS1}) and so the curves are smooth.

We also present the radius of the solutions defined as follows:
\begin{equation}
R = {\int r E(r) r^2 dr\over \int E(r) r^2 dr},
\label{RE}
\end{equation}
where $E(r)$ is the radial energy density.

Looking at the figures we see that, as $B$ gets very large, the normalized 
energy converges to a finite value when $p$ is even, but that it slowly 
diverges when $p$
is odd. This was also observed by Battye and Sutcliffe \cite{BS} in the case 
of $p=1$. When $m=4.131$ (Fig. \ref{figch}) the curve for $p=3$ crosses the 
value of $En=1$
at $B\approx 365$. For smaller value of $m$, the curves cross this
bound state threshold for much larger values of $B$. This is not surprising,
as for odd $p$, the mass term adds a non vanishing contribution to the energy 
density from the region where $f=\pi$  and so to make a large shell 
configuration one needs to put an increasing
amount of energy inside the shell. When $B$ is large, this becomes 
energetically too 
expensive and the configuration is not a bound state anymore.

We also see that, for a given parity of $p$, the energy at a given value of 
$B$ decreases as we increase $p$ by a multiple of 2, {\it i.e.} 
$En_p(B) > En_{p+2}(B))$. 

The plots of the radius also show that when $p$ is odd the shell is smaller, 
but otherwise, for a given parity of $p$, the radius increases with $p$.
On the other hand, for fixed values of $p$ and $B$, the radius decreases when
the mass $m$ increases. This is exactly what one expects for odd values of $p$
as the energy density inside the shell is non zero, but it is also true for 
even $p$, {\it i.e.} the radius of the shell decreases with the increase of 
the mass.

\begin{figure}[htb]
\unitlength1cm \hfil
\begin{picture}(15,9)
 \epsfxsize=8cm \put(0,0){\epsffile{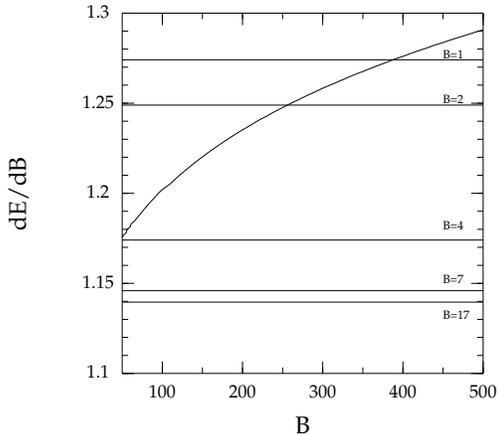}}
\end{picture}
\caption{\label{figDeB}
Derivarive of the normalized Energy with respect to $B$ and the normalized 
energy of $B=2$, $B=4$, $B=7$ and $B=17$ for $m=0.362$ and $p=1$.}
\end{figure}

One property worth investigating for these low energy configurations is their
ability to decay into two or more shells of smaller baryon charges. To do 
this we have computed the derivative of the energy with 
respect to 
$B$ and compared thus obtained values with the energy per baryon of some
small $B$ configurations of low energy (typically $B=2$, $B=4$, $B=7$ and 
$B=17$). When the value of the derivative is larger than the energy per baryon 
of 
other configurations, it implies that the larger configuration can decay into 
two shells. In Fig. \ref{figDeB} we see that, when $m=0.362$ and $p=1$,
all large shells ($B>50$) can decay into shells with $B=17$, 
$B=7$ and $B=4$ but can only decay into a $B=2$ if $B>250$ and $B=1$ if
$B>380$. When $p$ is even but non zero, the normalized energy decreases as $B$ 
increases implying that the binding energy of the configurations increases with
$B$. This in turn implies that the configurations never decay into smaller 
shells.

We summarize our observation in Table \ref{tableDecay} where we have given the
threshold value corresponding to several decay modes. The threshold values 
are the values of $B$ above which the decay is always possible, but sometimes 
some configurations with lower values of $B$ (smaller than 32) can decay in 
such a mode too (but only for very special values of $B$).
The values given with a `$\ge$` refer to the values obtained by comparing the 
energy of the configuration directly (low $B$) instead of using the
derivative of the energy.
 
\begin{table}
\begin{tabular}{|l|l|l|l|l|l|}
\hline
\bf m=0.362 &    &          &          &           &       \\ \hline
p   &$B=1$     & $B=2$    & $B=4$    & $B=7$     & $B=17$ \\ \hline
\hline
0   & $> 95$   & $> 70$   & $\ge 26$ & $\ge 18$  & $\ge 20$   \\ \hline
1   & $> 380$  & $> 250$  & $\ge 50$ & $\ge 28$  & $\ge 27$   \\ \hline
3   & -        & -        & -        & $>410$    & $> 100$  \\ \hline
5   & -        & -        & -        & -         & $> 390$  \\ \hline
\bf m=1.300 &       &          &          &           &       \\ \hline
p   &$B=1$     & $B=2$    & $B=4$    & $B=7$     & $B=17$ \\ \hline
\hline
0   & $\ge 18$ & $\ge 18$ & $\ge 8$  & $\ge 7$   & $\ge 17$   \\ \hline
1   & $\ge 27$ & $\ge 25$ & $\ge 14$ & $\ge 8$   & $\ge 18$   \\ \hline
3   & -        & $>480$   & $>120$   & $\ge 45$  & $\ge 34$   \\ \hline
5   & -        & -        & $>350$   & $> 110$   & $\ge 33$   \\ \hline
\bf m=4.131 &       &          &          &           &       \\ \hline
p   &$B=1$     & $B=2$    & $B=4$    & $B=7$     & $B=17$ \\ \hline
\hline
0   & $\ge 18$ & $\ge 18$ & $\ge 8$  & $\ge 8$   & $\ge 18$   \\ \hline
1   & $\ge 18$ & $\ge 18$ & $\ge 8$  & $\ge 8$   & $\ge 18$   \\ \hline
3   & $>160$   & $>110$   & $\ge 38$ & $\ge 24$  & $\ge 24$  \\ \hline
5   & $>390$   & $>240$   & $> 75$   & $\ge 36$  & $\ge 29$  \\ \hline
\end{tabular}
\caption{\label{tableDecay}
Decay of low energy configurations into sub shells. Each column corresponds to 
a decay mode and the baryon charge given corresponds to the threshold from 
which the decay is always possible.}
\end{table}

\begin{figure}[htbp]
\unitlength1cm \hfil
\begin{picture}(15,13)
 \epsfxsize=8cm \put(0,8){\epsffile{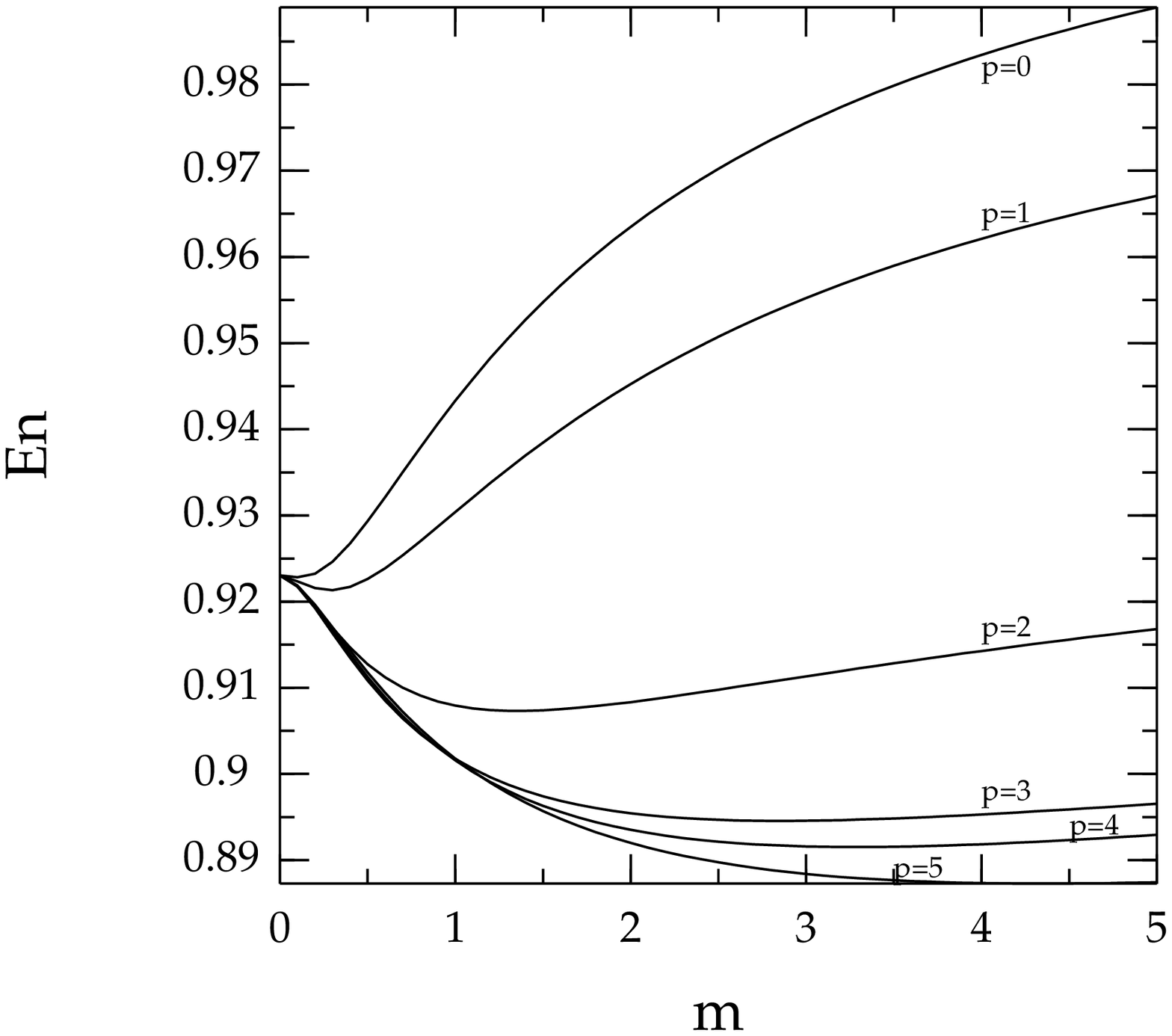}}
 \epsfxsize=8cm \put(7,8){\epsffile{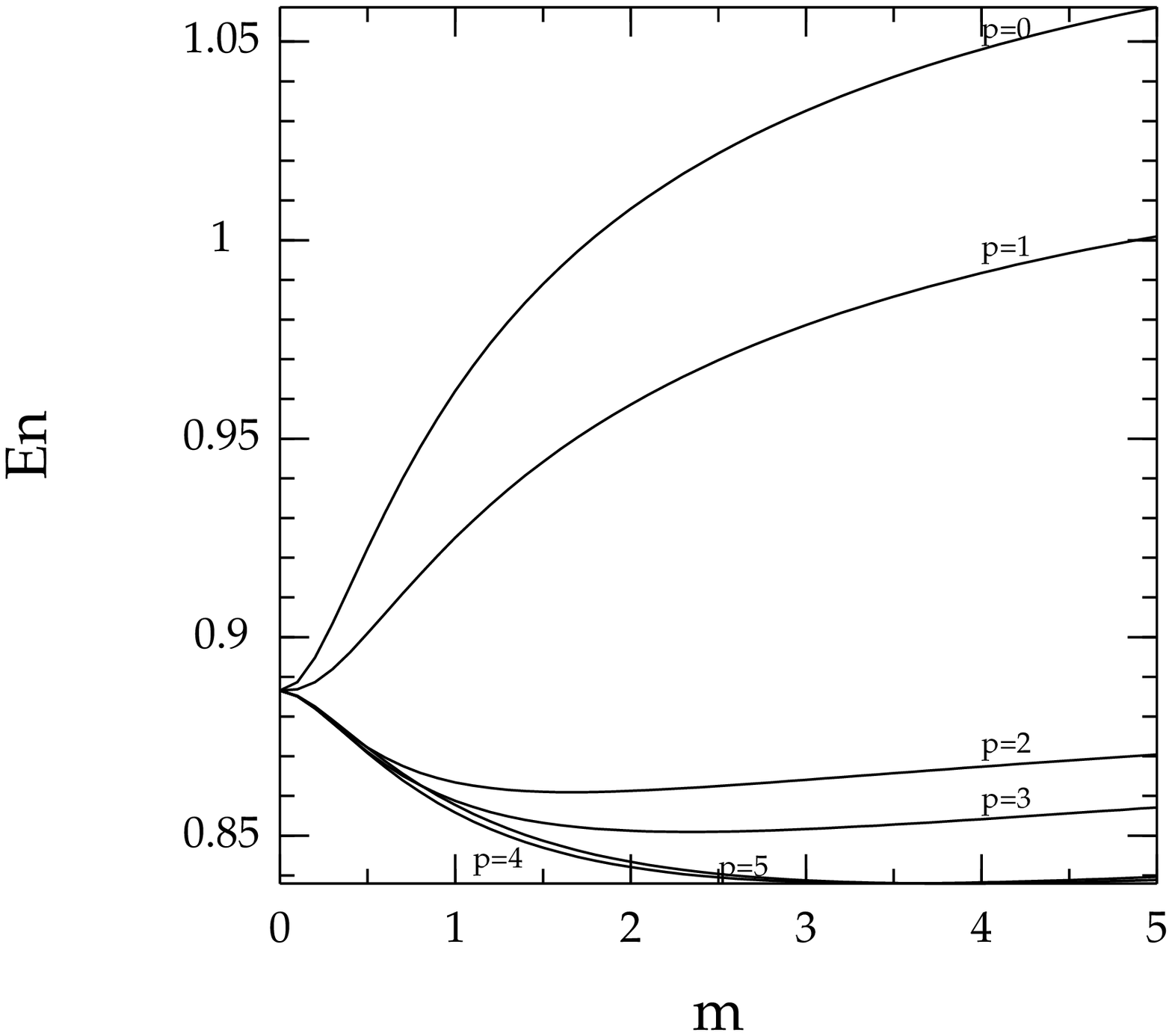}}
 \epsfxsize=8cm \put(0,1){\epsffile{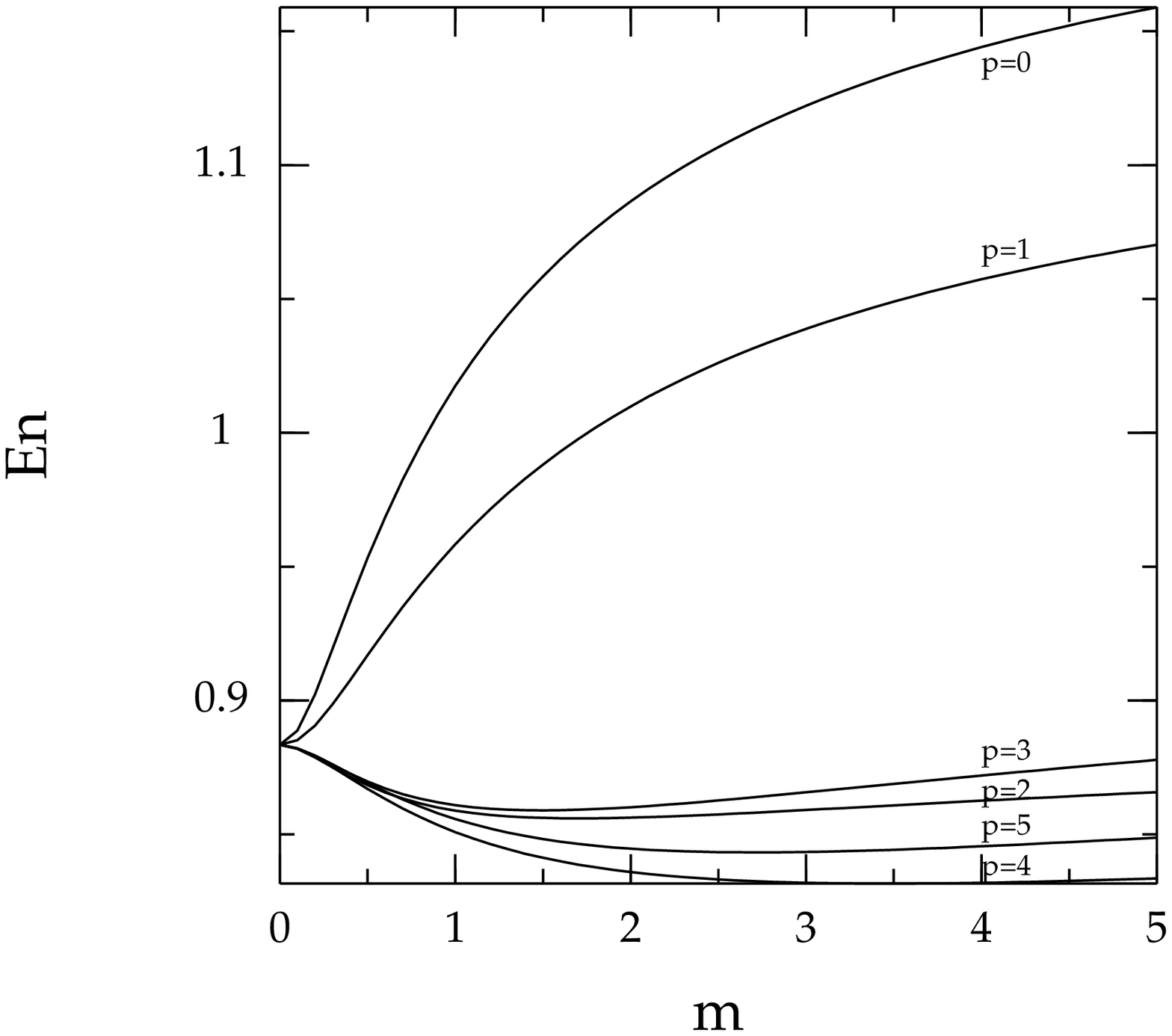}}
 \epsfxsize=8cm \put(7,1){\epsffile{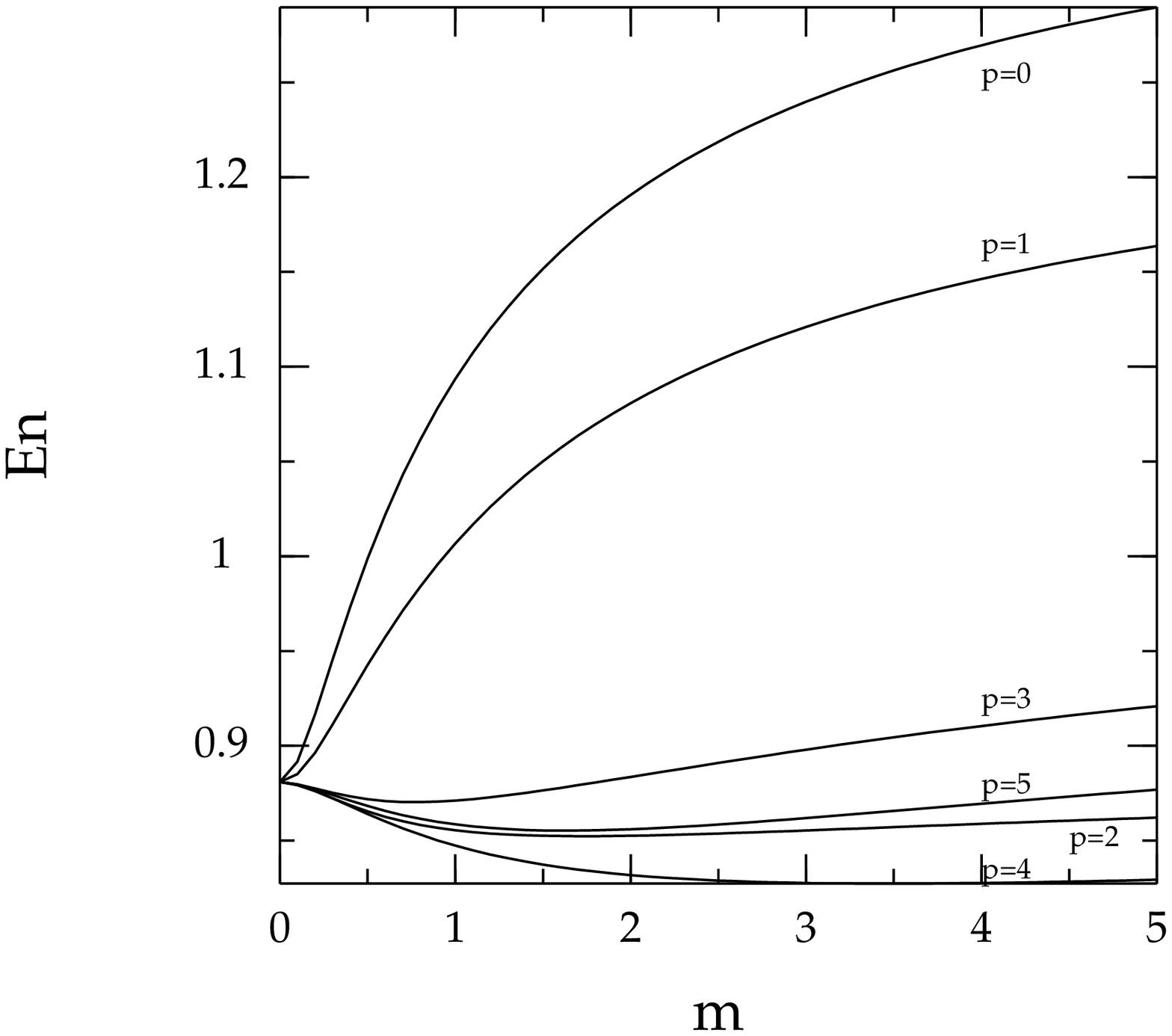}}
\put(3.0,8){a}
\put(10.0,8){b}
\put(3.0,1){c}
\put(10.0,1){d}
\end{picture}
\caption{\label{figMB}
Normalized energy $En$ as a function of $m$ for various values of 
$p$ for $B=4$ (a), $B=17$ (b), $B=40$ (c) and $B=100$ (d).}
\end{figure}

In Fig. \ref{figMB}, we present the normalized energy (\ref{NorEn})
as a function of $m$ for various values of $B$. We see that for $p=0$ and 
$p=1$ the energy increases rapidly with $m$ and that very quickly 
the configurations become 
unstable. When $p>1$, the normalized energy decreases for small 
$m$ when $m$ increases, then it reaches a minimum and finally it 
increases with $m$. The 
value of the mass for which the solution is the most bound depends on $p$ and 
on the baryon number.

In the next section we present an analytical description of the multiskyrmion
configurations which will explain some of the features we have observed 
numerically.

%
%
 %
%
%
%

%
%
%
%
%
%

\section{Approximate analytical treatment}
It was shown  in \cite{vk} that many properties of multiskyrmions, 
including
their classical mass, spatial distribution, moments of inertia, etc.,
can be described with a good accuracy using a relatively simple power-step (or
``inclined" step) approximation of the profile function. A similar approach 
was also used successfully to describe ``baby"-skyrmions in the $2+1$ 
dimensional Skyrme model \cite{PZ,KPZ,Weidig,IKZ}. 
This approximation also turns out  to be useful for the study of
the asymptotics of our massive multiskyrmion field configurations
 for large values of the baryon number $B$.

Let us consider first the large $r$ asymptotics of the profile function. 
Clearly, this asymptotic behaviour is governed by the 2nd derivative term and 
the mass term in the Lagrangian.
The Euler-Lagrange equation then becomes asymptotically 
\be
\label{asympt}
(r^2f')' = (2B + m^2r^2)f,
\ee
and so, if $m^2r^2 \gg 2B$, we have $2 rf'+r^2f'' = m^2r^2 f$, which has the
asymptotics
$f \sim exp(-mr)$. 

For the values of $B$ in the region of $r$ where $m^2r^2 < 2B$ the 
profile  $f$ has a power behaviour, and it is in this region that most of the
mass and most of the baryon density of the multiskyrmion is concentrated 
\cite{vk}, while the exponential
tail of the profile function gives only a small correction to all such
quantities and so can be neglected. Thus if we can neglect the $\sim m^2$ term 
on the right
side of (\ref{asympt}), we obtain the power law $f\sim r^{-\sqrt{2B}}$. 
As we shall see, the dominant range of $r$ is always such
that we can make  this approximation,  at least for pions and kaons.

Denoting $\phi=\cos f$ and taking for $g(k)=\delta(k-p)$ where $p$ is an 
integer, the energy of multiskyrmion can be written as
\bea
\label{mphi}
M&=&{1\over 3\pi}\int \biggl\{{1 \over (1-\phi^2)}
\bigl[r^2\phi'^2+2B(1-\phi^2)^2\bigr]\nonumber\\
&+&\biggl[2B\phi'^2+ \Ic {(1-\phi^2)^2\over r^2}\biggr] + 
2 m^2 \Psi_p(\phi) r^2\biggr\}dr,
\eea
with $\phi$ changing from $-1$ at $r=0$ to $1$ at $r\to\infty$.
The first part of (\ref{mphi}) is the second order term contribution while
the second term is due to the Skyrme term. 
Note that at fixed $r=r_0$ the $4$-th order term 
is exactly proportional to a $1$-dimensional domain wall energy widely 
discussed  in the literature, see e.g. \cite{koz}. The function 
$\Psi_p(\phi)=(1-\cos(pf))/p\sp2$ can be written
explicitly for each $p$: 
$\Psi_1 = 1-\phi,\; \Psi_3=(1-\phi)(1+2\phi)\sp2/9 \leq \Psi_1$,
$\Psi_2 = (1-\phi\sp2)/2 \leq \Psi_1$, 
$\Psi_4=\phi\sp2(1-\phi\sp2)/2 \leq \Psi_2$, etc.
Also it can be shown that $\Psi_3 \leq \Psi_2$, $\Psi_4\leq \Psi_3$, and it 
follows immediately, 
for any $p$, that $\Psi_{4p}\leq \Psi_{3p} \leq \Psi_{2p}\leq \Psi_p $, etc.
The functions $\Psi_p$ and the whole mass term have different properties for 
odd and even $p$
and so, for this reason, these two cases will be considered separately.

It is possible to rewrite the second order term contribution in (\ref{mphi}) 
as:
\be
M^{(2)}={1\over 3\pi}\int \biggl\{{ r^2 \over (1-\phi^2)}
\biggl[\phi'- \sqrt{2B}(1-\phi^2)/r\biggr]^2 + 2r\sqrt{2B}\phi' 
\biggr\}dr,
\ee
and similarly for the $4$-th order Skyrme term. Next we observe that
if $\phi$ satisfies $\phi'=\sqrt{2B}
(1-\phi^2)/r$,  a large part of the integrand in $M^{(2)}$ vanishes.
Therefore, it is natural to consider a function $\phi$ which satisfies the 
following differential equation \cite{vk}:
\be
\label{diffeq} \phi'= {b \over 2r}(1-\phi^2),
\ee
where $b$ is a constant.  A solution of this equation, which satisfies the 
boundary conditions 
$\phi(0)=-1$ and $\phi(\infty )=1$, is given by:
\be
\label{phi}
\phi (r,r_0,b) = \frac{(r/r_0)^b-1}{(r/r_0)^b+1} 
\ee
where $r_0$ is the distance from the origin to the point where 
$\phi=0$ and at which the profile $f=\pi/2$. $r_0$ can be considered as the 
radius of
the multiskyrmion. Both $b$ and $r_0$ are arbitrary at this stage;
they will be determined later by means 
of the mass minimization procedure. Note that the radii of distributions of 
baryon number and of the mass of the multiskyrmion are close to $r_0$.
Let us point out that our parametrization (\ref{phi}) is very accurate as, in 
the Skyrme model
with the usual mass term, as shown in \cite{vk}, the masses and other 
characteristics 
of multiskyrmions are described by such a parametrization to within a few $\%$.

\subsection{Odd powers, $p=1,\,3,...$}
Consider first the case of $p=1$. Then, using (\ref{rmaenergy}) and 
(\ref{phi}) we find that the soliton mass is given by
\bea 
\label{mphi1}
M(B,b) = {1 \over 3\pi}\int \biggl\{\biggl(\frac{b^2}{4}+2B\biggr)(1-\phi^2)
& + &\biggl( \Ic + {Bb\sp2\over 2}\biggr)\frac{(1-\phi^2)^2}{r^2}+ \nonumber\\
& + & 2 r^2 m^2(1-\phi) \biggr\} dr, 
\eea
where $\phi$ is given by (\ref{phi}) and where we should take
 $m=0.362$ for the pion case and $m=1.30 $ for kaons, etc.

Given the form of $\phi$ the integration over $r$ can now be performed using 
the well known expressions for the Euler-type integrals, e.g. 
\be
\label{eul}
\int_0^\infty \frac{dr}{1+(r/r_0)^b}=\frac{\pi r_0}{b\, \sin(\pi/b)},
\ee
if $b>1$, 
and, more generally \cite{vk},
\be
\label{eul1}
\int_0^\infty \frac{(r/r_0)^cdr}{\beta+(r/r_0)^b}=\beta^{(1+c-b)/b}
\frac{\pi r_0}{b\, \sin[\pi(1+c)/b]},\ee
with $\beta > 0, \;b > 1+c,\; c > -1$ .
Differentiating with respect to $\beta$ allows us to get the integrals with 
any power of $1+(r/r_0)^b$ in the denominator.
Thus we can derive the following expressions for the integrals of $\phi$ 
given by 
(\ref{phi}):
\be
\label{int}
\int (1-\phi^2)\, dr = \frac{4\pi r_0}{b^2 \sin(\pi/b)}, \qquad
\int \frac{(1-\phi^2)^2}{r^2} \,dr \;=\;
\frac{8\pi(1-1/b^2)}{3r_0b^2\sin(\pi/b)},
\ee
and other examples useful for the calculation of the mass term,
\be
\label{int1}
\int (1-\phi)\,r^2 dr = \frac{2\pi r_0^3}{b \,sin(3\pi/b)}, \quad
\int \phi^2(1-\phi^2)r^2 \,
  dr =\;\frac{4\pi r_0^3\bigl(1+18/b^2\bigr)}{b^2\sin(3\pi/b)}.
\ee
The expressions (\ref{int},\ref{int1}) allow us to obtain the mass of the 
multiskyrmion field in a simple 
analytical form as a function of the parameters $b$ and $r_0$ (in units 
$3\pi^2 F_\pi/e$):
\be
\label{Mabd}
  M(B,r_0,b) = \al (B,b) r_0 + \beta (B,b)/r_0 + \dl (b) \,r_o^3 .
\ee
where
\bea
\al &=& (b^2+8B)/\bigl(3b^2\sin(\pi/b)\bigr),\nonumber\\  
\beta&=&4(Bb^2 +2{\cal I})(1-1/b^2)/\bigl(9b^2\sin(\pi/b)\bigr),\nonumber\\ 
\dl &=&4m^2/\bigl(3b\,\sin(3\pi/b)\bigr).
\eea
The mass term contribution is proportional to the volume of the multiskyrmion, 
$\sim r_0^3$, as expected on general grounds,  
multiplied by the corresponding flavour content of the skyrmion 
\footnote{There is a difference of principle between the pion and kaon (or 
other flavoured meson) fields included into the Lagrangian. The mass of 
flavoured mesons enters into the classical soliton mass multiplied
by a corresponding flavour content which is always smaller than $1$ and 
even smaller than $0.5$ for a rigid or soft rotator quantization scheme. 
So, when we take the masses $m\simeq 1.30$ for the strangeness or 
$m\simeq 4.13$ for the charm we establish the scale of the mass for these 
flavours, not more than that.}.
Next we minimize (\ref{Mabd}) with respect to $r_0$ and 
obtain, in a simple form, the precise minimal value of the mass 
\be
\label{M5}
M(B,b)={2r_0^{min} \over 3}\bigl(\sqrt{\alpha^2+12\delta \beta}+2\alpha\bigr)
\ee
where the value of $r_0$ is given by:
\be
\label{r0}
r_0^{min}(B,b) =\biggl[\frac{\sqrt{\alpha^2 +12\beta \dl }-\alpha}{6\dl}
\biggr]^{1/2}. \ee
Eq. (\ref{Mabd}, \ref{M5})  give the upper bounds for the mass of the 
multiskyrmion state,
because they are calculated for the profile (\ref{phi}) which is different 
from the true profile to be obtained by the true minimization of the energy 
functional (\ref{genen}) with the mass term included. 
At large values of $B$ the power $b$ is also large, $b\sim \sqrt{B}$, as we 
shall see,
and $\al \simeq \bigl(b+8B/b\bigr)/(3\pi)$, 
$\beta \simeq 4\bigr(Bb+2 \Ic/b\bigr)/(9\pi)$, $\dl \simeq 4m^2/(9\pi)$.

The structure of (\ref{Mabd}) remains the same for values of $p$ 
different from
$1$, except for the case of even $p$ which will be  considered separately. For 
$p=3,\,5,\, $ etc. 
one must perform the substitution $\dl \to \dl /p^2$, and the volume 
contribution is reduced by a factor $1/p^2$.
The energy (\ref{M5}) can be simplified and analyzed in two different cases, 
small $m$ or
$\dl$, when $12\dl \beta \ll \alpha^2$ (which we will call in what follows 
the small mass
approximation, or SMA), and in the case of large $m$ or large $B$ when 
$12 \dl \beta \gg \alpha^2$, which we will call the large mass approximation, 
or LMA. 
Note, that at large $B$-numbers,
$\al \sim \sqrt{B}$ and $\beta \sim B\sqrt{B}$; therefore, when $B$ is large 
enough, the latter inequality  can always be satisfied: it reads then, 
approximately, $m^2\sqrt{B} \gg 1$.

Let us consider first the latter case of large $\beta \delta $ (the 
LMA case).  Now we can neglect the term
$\sim \al^2$ in the square root of (\ref{M5},\ref{r0}), and obtain
\be
\label{M6}
M(B,b) \simeq {4\over 3^{3/4}}\biggl(\beta^3 \dl\biggr)^{1/4}
\biggl[1+{\alpha \over 4}\biggl({3 \over \beta \dl}\biggr)^{1/2}\biggr]
\ee
and
\be
\label{r01}
r_0^{min} \simeq \biggl({\beta\over 3\dl}\biggr)^{1/4}
\biggl[1-{\alpha \over 4}\biggl({1 \over 3\beta \dl}\biggr)^{1/2}\biggr]
\ee
It is clear that the minimum value of the mass is reached at the minimum of 
$\beta$ ($\dl$ does not depend on $b$ when $b$ is large, and the correction 
term in the square bracket
has little influence on the position of the minimum),
which is equal to $\beta^{min} = 8\sqrt{2B \Ic}/(9\pi)$ at 
$b=\sqrt{2\Ic/B}$. Then
\be
\label{M7}
M(B) \simeq {16\over 9\pi}\biggl({2\over 3}\biggr)^{3/4} m^{1/2}
 \bigr(2B{\cal I}\bigr)^{3/8}
\Biggl[1 + \frac{3\sqrt{3}\,(\Ic +4B\sp2)}{8\sqrt{2}\,m (2\Ic B)\sp{3/4}} 
\Biggr]
\ee
Since $\Ic \sim B^2$ (strictly, $\Ic \geq B^2$ \cite{HMS}), we establish the 
following scaling 
law at large $B$: 
$M(B)\sim B\sp{9/8}m\sp{1/2}, \; r(B)\sim B\sp{3/8}/m\sp{1/2} $.
Numerically, we have for $p=1$ ($\Ic =1.28B^2$ in these estimates)
\be
\label{num1}
{M \over B}(p=1) \simeq 0.59395 \sqrt{m} B^{1/8} 
\Biggl( 1 \,+ {1.1982 \over mB\sp{1/4}}\Biggr)
\ee
For other odd $p$, dividing the volume contribution to the mass term by $p^2$, 
we obtain

\be
\label{num3}
{M \over B}(p) \simeq {0.59395\over \sqrt{p}} \sqrt{m} B^{1/8} 
\Biggl( 1 \,+ p{1.1982 \over m\,B\sp{1/4}}\Biggr).
\ee

\begin{table}
\begin{tabular}{|l|l|l|l|l|}
\hline
B & p=1 (num.)    & p=1 (SMA)    &  p=3 (num.) & p=3 (SMA)  \\ \hline
1   & 1.2740 & ---      & 1.2576 & --- \\
40  & 1.1519 & 1.1497 & 1.0990 & 1.0935  \\
100 & 1.1734 & 1.1760 & 1.0991 & 1.0982   \\
200 & 1.1973 & 1.1956 & 1.1023 & 1.1034  \\
300 & 1.2144 & 1.2037 & 1.1053 & 1.1073  \\
400 & 1.2279 & 1.2057 & 1.1080 & 1.1106  \\
500 & 1.2392 & 1.2038 & 1.1104 & 1.1133  \\
\hline
\end{tabular}
\caption{\label{tabpi} Energy per baryon for $m=m_\pi=0.362$. The analytical 
calculations 
are made in the small mass approximation (SMA) according to (\ref{delm}).}
\end{table}

\begin{table}
\begin{tabular}{|l|l|l|l|l|l|}
\hline
B & p=1 (num.)        & p=1 (LMA)&p=3 (num.) & p=3 (LMA) &p=3 (SMA) \\ \hline
1   & 1.4860 & ---      & 1.3842 & --- & ---  \\
40  & 1.4525 & 1.4675 & 1.1893 & 1.3018 & 1.1701 \\
100 & 1.5375 & 1.5553 & 1.2097 & 1.3032 & 1.1968 \\
200 & 1.6181 & 1.6351 & 1.2360 & 1.3157 & 1.2056 \\
300 & 1.6714 & 1.6875 & 1.2554 & 1.3276 & 1.1982 \\
400 & 1.7120 & 1.7273 & 1.2710 & 1.3381 & 1.1820 \\
500 & 1.7450 & 1.7596 & 1.2841 & 1.3474 & 1.1603 \\
\hline
\end{tabular}
\caption{\label{tabs} Energy per baryon for $ m=m_s=1.300.$  The analytical 
estimates are 
made according to (\ref{num3}) (LMA) and, for $p=3$, also in SMA.}
\end{table}
The absolute lower bound for the energy which follows from (\ref{num3}) 
obviously does not depend on $p$.
These estimates can be improved further: the $O(\al^2)$ 
terms in the expansion of the square root in (\ref{M5}) can be included; the 
surface contributions to the mass term, besides
the volume-like one, can be calculated (this may be important for higher 
$p$ since the volume
contribution decreases like $\sim 1/p^2$); the shift in the position of 
$b^{min}$ could also be taken into account.

As the ratio $M(B)/B \sim B^{1/8}$, we conclude that, for large $B$, the shell 
configurations will not form a bound state. This confirms what we have 
observed numerically. The second order term of the initial Lagrangian makes a 
small contribution in the large mass
regime; thus the Skyrme term and the mass term approximately balance each 
other, and the mass term gives $\sim 1/4$ of the total mass,
by the Derrick theorem. The difference between the cases $p=3,\, 5,...$ etc.
and $p=1$ resides in the fact  that for larger $p$ this
``large mass term regime" is reached at higher values of the baryon number.
One of the properties of multiskyrmions in this regime is that the average 
energy density does not depend on $B$, $\rho_M \sim m^2$, or, in
ordinary units, $\rho_M \sim \mu_\pi^2 F_\pi^2$ (which does not depend on the 
Skyrme parameter $e$). The energy density in the shell can be estimated as 
well; we get
$\rho_{shell} \sim \sqrt{B} \mu_\pi\sp2 F_\pi\sp2$ which grows when the 
baryon number increases.
And, as it has been previously discussed in the literature, the transition to 
other types of classical
configurations,  like the skyrmion crystals, may become possible at high 
values of $B$.

\begin{table}
\begin{tabular}{|l|l|l|l|l|}
\hline
B & p=1 (num.)    & p=1 (LMA)    &  p=3 (num.) & p=3 (LMA)  \\ \hline
1   & 2.0558 & ---      &1.7370  & --- \\
40  & 2.1778 & 2.1352 & 1.5160 & 1.4877  \\
100 & 2.3619 & 2.3436 & 1.5839 & 1.5805  \\
200 & 2.5304 & 2.5216 & 1.6589 & 1.6643  \\
300 & 2.6398 & 2.6344 & 1.7111 & 1.7191  \\
400 & 2.7222 & 2.7185 & 1.7516 & 1.7607  \\
500 & 2.7888 & 2.7861 & 1.7850 & 1.7945  \\
\hline
\end{tabular}
\caption{\label{tabc} Energy per baryon for $m=m_{ch}=4.131$. The analytical 
estimates are 
made in the large mass approximation, (\ref{num1},\ref{num3}).}
\end{table}

When the mass $m$ is small enough, as for the pion, the expansion in 
$12\beta \dl /\alpha^2$ can be
made, and one obtains the  reduction of the multiskyrmion size $r_0$:
\be
\label{r0sm}
r_0 \to r_0 - {3\dl \over 2\alpha} \biggl(\frac{\beta}{\alpha}\biggr)^{3/2}
 \simeq \sqrt{{2\over 3}}\Ic^{1/4}
  \biggl[1-\frac{2 m^2}{3} \Ic^{1/4}\eta_B\biggr], 
\ee
and the increase of its mass
\be
\label{delm}
\dl M = M_{m=0} \frac{\beta \dl}{2\alpha^2}
\biggl[1-\frac{9\beta \delta}{8\alpha^2} \biggr]
\simeq M_{m=0}  \frac{2m^2}{9} \Ic^{1/4}\eta_B
\biggl[1 - \frac{m^2}{2} \Ic^{1/4}\eta_B \biggr],
 \ee
$\eta_B = \sqrt{\Ic}/(2B +\sqrt{\Ic})$,
$M_{m=0} \simeq 4B\sqrt{2/3}(2+\xi_B)/(3\pi)$ \cite{vk}, and at large $B$,
$M_{m=0}/B \simeq 1.0851$, $\eta_B \simeq 0.3613$ if we take 
$\xi_B =\sqrt{\Ic/B^2}\simeq 1.13137$ - 
constant value, according to \cite{BS1}.
Here, we have used also the observation that, at large $B$, 
$b^{min}=2{\cal I}^{1/4}$ \cite{vk} and
$$\alpha \simeq {2\over 3\pi{\cal I}^{1/4}}\biggl(2B+\sqrt{{\cal I}}\biggr) 
\sim \sqrt{B}, 
\qquad \beta \simeq {4 \Ic^{1/4}\over 9\pi}\biggl(2B + \sqrt{{\cal I}}
\biggr) \sim B^{3/2}. $$
As expected, the size of the multiskyrmion state decreases
with increasing $m$ while its mass increases, and these changes become very 
large for very large $B$ and/or $m$. 

For $p$ different from $1$ the substitution $m^2 \to m^2/p^2$ should be made 
in (\ref{r0sm},\ref{delm}) and the following relation can then be obtained 
for any pair of odd $p$'s, $p_1$ and $p_2$:
\be
\label{ratio}
{M_B(p_1) - M_B(p_2)\over M_B} \simeq {2m^2\over 9} 
\biggl({1\over p_1^2} - {1\over p_2^2}\biggr) \Ic^{1/4}\eta_B
\ee
Numerically this works well for $p=1$ and $p=3$, see Table 
\ref{tabpi}-\ref{tabc};  for larger $p'$s
the agreement is less good but then, apparently, other contributions
to the mass term, besides the volume-like one, should be included.

In Tables \ref{tabpi}-\ref{tabc}, we present several values of the energy per 
baryon obtained 
from (\ref{delm}) (Table \ref{tabpi}) and (\ref{num1}) and (\ref{num3}) 
(Tables \ref{tabs},\ref{tabc}),  
and compare them  with the values obtained numerically. We see that our
analytical approximation works very well when the mass is small and 
$B$-numbers not too large (pions case, Table \ref{tabpi}),
or when it is large, as for the charm, but it does not work so well for 
intermediate values of the mass. It also works better for $p=1$ than for $p=3$.
The case $p=3$, presented in Table \ref{tabs}, is of special interest: the LMA 
improves when increasing the
baryon number but is still not as good as for $p=1$, whereas SMA becomes 
worst when $B$ increases and also is not perfect at small values of $B$. 
To improve it, the following terms in the expansion (\ref{delm}) 
should be included.

For not very large values of $m$ the structure of the multiskyrmion 
at large $B$
remains the same: it is given by the chiral symmetry broken phase inside a 
spherical wall where (on this spherical shell) the main contribution to the 
mass and topological 
charge is concentrated \cite{vk,BS1}. The value of the mass density inside
this wall is defined completely by the mass term with $1-\phi= 2$ and decreases
with increasing $B$ while the mass density of the shell itself is constant 
\cite{vk}.
The baryon number density distribution is quite similar, the only difference
being that inside the spherical wall it vanishes. 

If, for some physical reasons, we would use a distribution over $p$ in the 
Lagrangian, as discussed in Sections 2,3, the analytical expression for the 
multiskyrmion energy can be obtained quite analogously. Let us put 
$p=p_0+ \Delta p$, where $p_0=1,\;3$, etc., and $\Delta p$ is assumed to be 
small. Then, taking into account the changes in the
volume contribution to the mass term, we get instead of (\ref{num3}):
\bea
\label{nearpodd}
{M \over B}(p) \simeq {0.59395\over \sqrt{p_0}} \sqrt{m} B^{1/8}
   \Biggl[1 -{\Delta p\over 2p_0}-
   {(\Delta p)^2\over 16}\biggl( \pi^2 - {6\over p_0^2}\biggr) \Biggr] 
   \times \nonumber \\
    \times \Biggl[ 1 \,+ p_0\biggl(1+{\Delta p\over p_0} 
                   +{(\Delta p)^2\pi^2 \over 8}\biggr)
                  {1.1982 \over m\,B\sp{1/4}}\Biggr],
\eea
and the averaging over any distribution $g(p)$, as suggested by 
(\ref{maone},\ref{test},\ref{gauss}) can be easily performed.

It is also possible to consider, in a similar way, the case of small values of 
$p$, near $p=0$.
In this case we have $(1- \cos (pf))/p^2 \simeq f^2 (1-p^2f^2/12)/2 $, and 
$f\simeq \pi$ inside the multiskyrmions.
Evaluations similar to those at the beginning of this section show that  
the energy per unit $B$-number, in the large mass regime, is given by:
\be
\label{smallp}
{M\over B} \simeq 0.74441\sqrt{m}B^{1/8}\biggl(1-{p^2\pi^2\over 48}\biggr)
\Biggl[1+{0.7628\over mB^{1/4}}\biggl(1+{p^2\pi^2\over 24}\biggr)\Biggr].
\ee
Obviously, at large enough value of the mass, this is somewhat greater than 
the energy given by (\ref{num1}).
The expression (\ref{nearpodd}) can be compared with our numerical results for 
$p=0$ presented in 
Fig. \ref{figch} and Fig. \ref{figst}. We note the agreement to within 
an accuracy of about $(5-8)\%$ for the largest 
values of $m$ and $B$. The integration
over any distribution in $p,\;g(p)$, near $p=0$, as presented in 
(\ref{maone},\ref{test}), can also be easily made.
\subsection{Even power $p=2,\,4,...$}
In the case of even $p$ \ie\ $p=2,\,4...$ the volume contribution to the 
energy density is reduced because $1 - \cos (pf) \simeq 0$ inside the 
multiskyrmion where the profile $f \simeq \pi$.
Due to the dependence of the mass term on the parameter $b$ and due to the 
connection between $r_0$ and $b$ this case is very different from the case of 
$p=1,\,3...$. However, we can still write
\be
\label{Mabdd}
M(B,r_0,b)\simeq \alpha(B,b) r_0+ {\beta(B,b)\over r_0} 
+ {\dl '(b)r_0^3\over b} . \ee
This expression coincides with  (\ref{Mabd}), except for the mass 
term where an additional factor
$1/b$ appears and $\dl'$ is different from $\dl$. For $p=2$ we have 
$\dl '(p=2)=4m^2/(b\, \sin(3\pi/b)) \simeq 4m^2/(3\pi)$ at large $b$ while for 
$p=4$ an additional 
small factor appears as 
$\dl '(p=4)=4m^2(1+18/b^2)/(3b\,\sin(3\pi/b)) \simeq 4m^2/(9\pi) $. 
It is not easy to find the general expression for larger $p$; \ie\  the 
expression for $\dl '$ which decreases with increasing 
$p$ (see the discussion after (\ref{mphi})).

In general, we proceed as for odd values of $p$ and after minimizing with 
respect to $r_0$ we obtain
\be
\label{M5e}
M(B,b)={2r_0^{min} \over 3}\bigl(\sqrt{\alpha^2+12\dl ' \beta/b}+
2\alpha\bigr)
\ee
and the value of $r_0$
\be
\label{r0e}
r_0^{min}(B,b) =\biggl[\frac{\sqrt{\alpha^2 +12\dl ' \beta/b} 
-\alpha}{6\dl '/b}
\biggr]^{1/2}. \ee
The main difference from the previous case is that, at large values of $B$, 
the quantities
$\alpha^2 \sim B$ and $\dl '\beta/b \sim B$, \ie\ they are of the same order 
of magnitude, since
$b\sim \sqrt{B}$. 
At large enough $b$ or $B$ we have $12\beta \dl '/(b\,\alpha^2) \simeq 0.35$ 
for pions,
$4.5$ for kaons and $\sim 46$ for charm.

Let us discuss first the large mass case when we can take
\be
\sqrt{\al^2+12\dl' \beta/b} \simeq 2\sqrt{3\dl '\beta/b}
\ee
and
\be
M(B,b) \simeq {4\over 3\sp{3/4}}\biggl({\beta^3\dl '\over b}\biggr)^{1/4}
 \biggl[1 +
{\alpha \over 4}\biggl({3b \over \beta\dl '}\biggr)\sp{1/2}\biggr].
\ee
The minimum is reached at $b\simeq 2\sqrt{\Ic/B}$ and we have, recalling that 
at large
$B$ and $p=2$, $\dl ' \simeq 4m^2/(3\pi)$,
\be
\label{Bp2}
 M(B,p=2) \simeq B m^{1/2} {16 \xi_B\sp{1/2}\over 3\sqrt{3}\,\pi\, 6\sp{1/4}}
\Biggl[1 +
 \frac{3 \sqrt{3}\,(\xi_B+2/\xi_B)}{4\sqrt{2}\,m}\Biggr], 
\ee
at $r_0 \sim \sqrt{B/m}$. Note that $\xi_B = \sqrt{\Ic/B\sp2}$ and, at large 
$B$, it is constant within the rational map approximation.

Numerically, for $p=2$, we obtain from (\ref{Bp2}) 
\be
\label{num2}
{M\over B}(p=2) \simeq 0.66612\, \sqrt{m}\Biggl(1 + {0.8877\over m} \Biggr)
\ee
and for $p=4$
\be
\label{num4}
{M\over B}(p=4) \simeq 0.50614\,  \sqrt{m}\Biggl(1 + {1.5375\over m} \Biggr).
\ee

In Table \ref{tabeven}, we present a few values of the asymptotic energy 
obtained from (\ref{Mmt}), SMA, and from (\ref{num2}) and (\ref{num4}) in LMA
and compare them with the values obtained numerically for $B=500$. 
In our calculations, for large $B$, 
we have again used the value
$\xi_B =1.13137$, \cite{BS1},  and $\dl' \simeq 4m^2/(9\pi)$ for $p=4$.
There is a good agreement between our numerical results and our analytical
approximation values when the mass is small (pions), or large (charm scale), 
and not so good for the intermediate value $m=m_{st}$ where we give 
both the SMA and LMA results.
Our approximation also works better for $p=2$ than for $p=4$. Nevertheless,
analytical approximations work better for odd $p$ at large $m$ and $B$, than 
for the even ones. 
It is possible to improve the analytical estimates although,
for even $p$, the estimate of the preasymptotic contributions to the energy 
appears to be technically harder to obtain than for odd $p$.

\begin{table}
\begin{tabular}{|l|l|l|l|l|}
\hline
m & p=2 (num. B=500) & p=2 (approx) &  p=4 (num. B=500) & p=4 (approx) \\ 
\hline
0.362  & 1.0879 & 1.1016 (SMA) & 1.0881 & 1.0908 (SMA) \\
\hline
1.300  & 1.2186 & 1.2048 (SMA) & 1.1362 & 1.1475 (SMA) \\
       &        & 1.2781 (LMA) &        & 1.2596 (LMA) \\
\hline
4.131 & 1.6236 & 1.6448 (LMA) & 1.3519 & 1.4116 (LMA) \\
\hline
\end{tabular}
\caption{\label{tabeven} Asymptotic values of the energy per baryon for $p=2$ and $p=4$. The 
analytical calculations correspond to (\ref{Mmt}), SMA, 
and (\ref{num2},\ref{num4}) in LMA.}
\end{table}

So, for $p=2,\,4,...$ etc and for large meson masses the multiskyrmion mass is 
proportional to the baryon number, and the average (volume) mass density 
decreases as $\sim 1/\sqrt{B}$. At large $B$ the thickness, or width of the 
shell is given by $W \sim 1/\sqrt{m}$ - \ie\ it does not depend on the 
$B$-number and the mass density in the shell is 
constant, $\rho_{shell} \sim \mu_\pi\sp2 F_\pi\sp2 $, in contradistinction to 
the case of odd $p$ where it grows with $B$ \footnote{In this respect there 
is a direct analogy with the $(2+1)-D$ model \cite{PZ,Weidig,IKZ}, where the 
surface energy density of the rings,
representing states of lowest energy, and their width, do not depend on the
topological number when this number is large.}.

When the meson mass is small, as for pions, we can perform the following 
expansion
\be
\label{expan}
\sqrt{\al^2 + 12 \dl'\beta/b} \simeq \al + 6\dl '\beta/(b\al) 
+...
\ee
The main contribution to the mass is then $M_0 = 2\sqrt{\al\beta}$ at 
$r_0=\sqrt{\beta/\al}$, 
and the minimum is reached at $b^{min} \simeq 2 \Ic^{1/4}$, as in the 
massless case \cite{vk}.

As a result we obtain the following expression  for the mass of the 
multiskyrmion, given the mass term in the Lagrangian, to first order in 
$\dl '$:
\bea
\label{Mmt}
M(B) &\simeq & M_{m=0} \biggl(1 + \frac{\dl'\beta}{2b\al^2}\biggr)\nonumber \\
     &\simeq &
{4B\over 3\pi}\sqrt{{2\over 3}} \biggl(2+\sqrt{\Ic /B^2}\biggr)
\biggl[1 + \frac{\pi \dl'}{4}\eta_B
\biggl(1-\frac{9\pi \dl'}{16}\eta_B \biggr)\biggr].
\eea
At large values of $B$ the relative contribution of the $ m^2$ correction 
is constant (since $\dl'$ is constant at large $B$ and 
$\Ic/B^2 \to \mbox{const}$,
$\eta_B \to \mbox{const} = 0.3613$), in 
contradistinction to the case of odd values of $p$,
and the value of $M(B)/B$ from (\ref{Mmt}) is independent of $B$.
Note that the difference 
of the multiskyrmion masses, between the $p=2$ and $p=4$ cases, is given by:
\be
\label{M24}
{M(B,p=2)-M(B,p=4)\over M(B, m=0)} \simeq {2m^2\over 9}\eta_B 
  \bigl(1 - m^2\eta_B\bigr) .
\ee
For the pion mass $m^2\simeq 0.13$, and (\ref{M24}) gives the value 
$\sim 0.01$, 
as shown in Table \ref{tab24}, in agreement with the numerical data, for 
greater 
$m$ the agreement is not so good, since the case of $p=4$ is more difficult to
describe analytically. 

For any even $p$ the mass term gives a contribution to the multiskyrmion mass 
which is 
constant at large baryon numbers (relatively), and which decreases with  
increasing 
$p$, as then $\dl'$ decreases. This is in agreement with the numerical results 
presented in the previous section.

The radius of the multiskyrmion state, to first order in the mass term, can 
also be rewritten as
\be
\label{r0m}
r_0 \simeq r_{0,m=0}\biggl(1 - \dl '{3\beta\over 2\al^2b} \biggr)
\simeq r_{0,m=0}\biggl(1 - {3\pi\over 4}\dl'\eta_B \biggr).
\ee
Since $\dl'$ decreases with increasing $p$, the radius of the multiskyrmion 
increases, in good agreement with
the numerical results of the previous section.
From (\ref{Mmt}) and (\ref{r0m}) we have also that
\be
\label{relat}
\frac{r_B(p_2) - r_B(p_1)}{r_B(p_1)}  \simeq -3 \frac{M_B(p_2) - 
M_B(p_1)}{M_B(p_1)}
\ee
which is verified to a good accuracy  for $B$ larger than $\sim 10$.

\begin{table}
\begin{tabular}{|l|l|l|l|l|}
\hline
m &  B=100 &  B=400 & (approx) \\ 
\hline
$0.362$ & $0.00989 $         & $0.00987$ & $0.0100$ (SMA) \\
$1.300$ & $0.07602$          & $0.07615$ & $0.0528$  (SMA) \\
$4.131$ & $0.25053$         & $0.25089$ & $0.2149$ (LMA)   \\
\hline
\end{tabular}
\caption{\label{tab24} $(M(B,p=2)-M(B,p=4))/M(B, m=0)$ for numerical solutions 
(column 2 and 3) and according to SMA (\ref{Mmt}) and LMA 
(\ref{num2},\ref{num4}), column 4. }
\end{table}

To summarize, in the case of even $p$, \ie\  $p=2,\,4...$, the multiskyrmions have the 
structure of empty shells; the mass and $B$-number densities are concentrated 
in the envelopes of these shells and the energy per unit $B$
decreases with increasing $B$, assymptotically approaching a constant value.
\subsection{Even $p$; `the inclined step approximation'}
A natural question then arises: to what extent the structure of multiskyrmions 
and their properties
depend on the parametrization we have used. Of course, we have 
to satisfy the boundary conditions on the profile function: 
$f(0)=\pi$ and $f(\infty) = 0$
and the function should minimize the value of the mass (\ref{mphi}). However,
the profile $f$ could have been decreasing according to a law which is 
different from (\ref{phi}), thus giving us
 different mass and $B$-number distributions. But it is just the property 
of the Lagrangian  (\ref{mphi}) that produces the above-mentioned bubble 
structure as this structure leads to a  low value
of the mass. Another, perhaps the simplest,  example of a description that we 
can make is provided by the `toy'  model of 
``the inclined step" type \cite{vk}. Such an approximation is cruder than 
``the power step"
considered previously. However, it has the advantage that the calculations 
can be made for arbitrary $p$. Hence, we shall mention it here and compare its 
results with what we have obtained before.

Let $W$ be the width of the step, and $r_0$ - the radius of the 
multiskyrmion state, defined by the value of $r$ at which the profile 
$f =\pi/2$. 
Then we can approximate the profile function by $f= \pi/2 - (r-r_o)\pi/W$
for $r_o-W/2 \leq r \leq r_o+W/2 $.
This approximation describes the usual domain wall energy 
(see, e.g. \cite{koz}) to within an accuracy of $\sim 9.5 \%$.

Next, we write the energy in terms of $W,\, r_0$ (recall that $W\sim r_0/b$ in 
terms of the
previous parametrization) and minimize it with respect to both these 
parameters thus 
finding the approximate value of the energy. The case of $p=1$ was considered 
previously \cite{vk}, and since
the case of other odd $p$ is similar, we restrict our attention here to the 
case of even $p$'s. 

Thus for an arbitrary even $p$ we have:
\be
\label{intw}
{1\over p^2}\int_{r_0-W/2}^{r_0+W/2} \bigl(1-\cos(pf)\bigr)r^2 
dr= {r_0^2\over p^2}W + {W^3\over 12p^2} +
{2W^3\over p^4\pi^2}.
\ee
The volume term $\sim r_0^3$ is absent, and since $W\ll r_0$ at large $B$, we 
retain the term
$\sim r_0^2W$ on the right hand side of (\ref{intw}) and omit other terms.
Then, for the classical mass of the multiskyrmion, we have (the second 
and 4-th order terms were presented in
\cite{vk}):
\be
\label{MRW}
M(B,r_0,W) \simeq {1\over 3\pi}\biggl[{\pi^2\over W}(r_0^2+B) +
W\biggl(B+{3{\cal I}\over 8r_0^2}\biggr)
 + m^2 {2 r_0^2W\over p^2}\biggr].
\ee
The minimization with respect of $r_0$ is straightforward and it gives us
\be
\label{MW}
M(B,W)\simeq {1 \over 3\pi} \biggl[\sqrt{3\Ic}\biggl({m^2W^2\over p^2} +
{\pi^2\over 2}\biggr)^{1/2} +  B\biggl( {\pi^2\over W} + W\biggr)\biggr]
\ee
while $\bigl(r_0^{min}\bigr)^2=p \sqrt{3\Ic}/
\bigl[4m\sqrt{1+p^2\pi^2/(2m^2W^2)}\bigr]$.

We can now consider the two opposite cases. In the case of a large mass, when
$2m^2W^2 \gg p^2\pi^2$, we can expand $(m^2W^2/p^2+\pi^2/2)^{1/2} 
\simeq mW/p+\pi^2p/(4mW)$ and obtain
\be
\label{MWW}
M(B,W)\simeq {1\over 3\pi} \biggl[W\biggl(B+\sqrt{3\Ic}m/p\biggr) 
+ {\pi^2\over W}
\biggl(B+{p\sqrt{3\Ic}\over 4m} \biggr) \biggr].
\ee
This gives us
\be
\label{Mstep}
M(B) \simeq {2B\over 3}\biggl(1+\sqrt{3}\xi_B{m\over p}\biggr)^{1/2}
\biggl(1+ {\sqrt{3}p\xi_B\over 4m}\biggr)^{1/2}
\ee
for $W^{min} = \pi \bigl[\bigl(1+\sqrt{3}\xi_Bp/(4m)\bigr)/
\bigl(1 + m\sqrt{3}\xi_B/(2p)\bigr)\bigr]^{1/2}$ where 
$\xi_B = \sqrt{\Ic/B^2}$.
At large $m$ the simple formula (\ref{Mstep}) provides an asymptotic 
(at large $B$) value of the static
energy per unit $B$ which, as is easily seen, is in a good agreement with our 
numerical results, \ie\
 for $m\simeq 4.13$ it gives $M/B \simeq 1.666$ for $p=2$ and 
$M/B \simeq 1.408$ for $p=4$ 
which agree with the numbers in Table \ref{tabeven} to within $(3-4)\%$.

On the other hand, the large mass limit cannot be assumed when $p$ is large 
and so we can consider only the small mass limit: $2 m^2W^2 \ll p^2\pi^2$.
In this case we can consider the $m^2$ dependent term in (\ref{MW})
as a perturbation, as it was done in \cite{vk} and in the section above. 
Now $W^{min}=\pi$ and
\be
\label{smW}
M(B)\simeq {1\over 3}\biggl(2B + \sqrt{3\Ic/2} + 
{m^2\over p^2}\sqrt{3\Ic/2}\biggr).
\ee
The radius $r_0$ is now given by \cite{vk}
\be
r_0^2 \simeq \bigl(3{\cal I}/8\bigr)^{1/2}
\ee
and so, $r_0\sim \sqrt{B}$, and the corrections to $r_0$ can be easily found, 
following the steps similar to those of \cite{vk}.
The difference of masses of the $p=2$ and $p=4$ cases is reproduced well; 
however, this approximation is too crude at higher values of $p$.

To conclude, we see that the results obtained within 
``the inclined step" approximation reproduce well the results of the 
preceeding subsection for even
$p$ and describe well the transition to higher values of $p$ where the small 
mass limit can be applied.
Further refinements and improvements of this analytical discussion are 
possible, e.g. subasymptotics
of the $B$-number dependence can be calculated, but we shall not do this here 
since our results 
are already in a good agreement with the numerical data of section 5, and 
as the asymptotic behaviour of the solitons mass also is well understood, 
(\ref{num1},\ref{num3},\ref{num2}), etc. 
\section{Conclusions}
We have investigated different possible forms of the mass term in the $SU(2)$ 
Skyrme model,
concentrating our attention on a class of terms involving a parameter $p$.  
We have
found that the case of even $p$, but $p\ne 0$,  
in the parametrization of the mass term, 
$\sim (1-\cos(pf))/p\sp2$ is of special interest. In this case the 
contribution to the
static mass proportional to the volume of multiskyrmion is absent, and the 
multiskyrmion states, at any
value of the chiral meson mass $m$ in the Lagrangian, are the spherical 
bubbles empty inside, their
energy and baryon number being concentrated on the surface of the bubble. The 
energy per  
$B$-number decreases with increasing $B$ and approaches a constant value,  
which is proportional to 
$\sqrt{m}$ when $m$ is large. Thus as $B$ increases the shell configurations 
become more bound. 

For odd values of $p$ and $p=0$, there is a volume-like
contribution to the static mass of multiskyrmions, within the rational map 
approximation, which 
grows faster than $B$. Thus it is responsible for the asymptotic behaviour of 
the multiskyrmion mass
$\sim B\sp{9/8}\sqrt{m}$ and makes the multiskyrmions unstable 
with respect to the decay into skyrmions of smaller values of $B$. 
These unbound configurations have  recently been observed
by Battye and Sutcliffe \cite{BS}, for $p=1$, for relatively low values of $B$ 
even for the pion mass. We have also observed them for the other odd values of 
$p$: $p=3$ and $p=5$. When the mass $m$ is small, the configurations decay is 
into a $B=17$ or $B=7$ shell only for very large values of $B$. As the mass
increases, the decay becomes possible for smaller values of $B$.
When $p$ is even but non null, the configuration can't decay into smaller 
shells.
Thus if we want our states to be bound states we should consider $p$ even, 
or a combination of terms with $p$ even.

Note that the fact that some rational map configurations do not form 
classically bound 
states suggests that the states of minimal energy for those values of $p$ and 
$B$ can be of 
a different form. They might correspond to embedded shells, or not have any 
shell structure at all. This needs to be investigated further but this can be 
done only by solving the 
full equations of the model.

The configurations we obtained for $p>1$ have lower energy than those 
considered within the standard mass term and, for this reason, they have a 
good chance to find realization in nature, not only in elementary particle and 
nuclear physics, but also in astrophysics and cosmology. Of course, what 
really happens for physical nuclei is unclear as our results are classical;
\ie\ to compare them with physical nuclei we would have had to compute 
quantum corrections, and this has not yet been done for nonzero modes.

Investigations of multiskyrmions could be extended also to variants of the 
model
where the higher derivative terms (six order, eight order, etc.) are included 
into the effective Lagrangian. The studies performed in \cite{FP} for the case 
of the 6-th order term and recently in \cite{LM} for some generalizations of 
the Skyrme model including an 
8-th order term in the chiral derivatives, have shown that topological 
structures
of minimal energy configurations are the same for these model extensions as in 
the original variant of the model, for values of baryon numbers not too large.
Therefore, most probably, the observations of bound large $B$ states made in 
the present paper for certain modifications of the mass term will be confirmed 
in such generalizations of the model, although this requires detailed studies.

\section{Acknowledgement}
The work of VBK has been supported, in part, by the grant RFBR 01-02-16615.


\begin{thebibliography}{99}
\bibitem{AN} G. Adkins and C. Nappi, Nucl. Phys. B233, 109 (1983)
\bibitem{G} G. Guadagnini, Nucl. Phys. B236, 35 (1984)
\bibitem{vk} V.B. Kopeliovich, JETP 93, 435 (2001); JETP Lett. 73, 587 (2001); 
J.Phys. G28, 103 (2002)
\bibitem{BS} R.A. Battye and P.M. Sutcliffe, hep-ph/0410157 (2004)
\bibitem{M} B. Moussalam, Ann. Phys. (N.Y.) 225, 264 (1993) ; F.Meier and 
H. Walliser, Phys. Rept. 289, 383 (1997)
\bibitem{GL} J. Gasser and H. Leutwyler, Ann. Phys. 158, 142 (1984);
Nucl. Phys. B250, 465 (1985)
\bibitem{HMS} C.J. Houghton, N.S. Manton and P.M. Sutcliffe, 
Nucl. Phys. B {510}, 507 (1998).
\bibitem{IPZ} T.A. Ioannidou, B. Piette and W. Zakrzewski, J.Math.Phys. 40, 
6223 (1999); T.A. Ioannidou, B. Piette and W. Zakrzewski, J.Math.Phys. 40, 
6353 (1999)
\bibitem{BS1} R.A. Battye and P.M. Sutcliffe, Rev. Math. Phys. 14, 29 (2002)
\bibitem{PZ} B. Piette and W. Zakrzewski, Nonlinearity 9, 897 (1996)
\bibitem{KPZ} A.E. Kudryavtsev, B.M.A.G. Piette, and W.J. Zakrzewski, 
Nonlinearity 11, 783 (1998)
\bibitem{Weidig} T. Weidig, hep-th/9811238, Nonlinearity 12, 1489 (1999); 
hep-th/9911056
\bibitem{IKZ} T. Ioannidou, V. Kopeliovich and W. Zakrzewski, JETP 95, 572 
(2002), hep-th/02032053
\bibitem{koz} Ya.B. Zeldovich, I.Yu. Kobzarev and L.B. Okun', 
Zh. Eksp. Teor. Fiz. 67, 3 (1974)
\bibitem{FP} I. Floratos and B. Piette, Phys. Rev. D64, 045009 (2001)
\bibitem{LM} J.-P. Longpre and L. Marleau, hep-ph/0502253
\end{thebibliography}
\end{document}